\documentclass[12pt]{article}

\usepackage[utf8]{inputenc}
\usepackage[utf8]{inputenc}

\usepackage{amssymb}
\usepackage{amsfonts}
\usepackage{graphicx}
\usepackage{circuitikz}
\usetikzlibrary{quantikz}

\usepackage[verbose]{placeins} 

\usepackage{braket}
\usepackage{tikz}
\usepackage[T1]{fontenc}
\usepackage{imakeidx}

\usepackage{xcolor,varwidth}

\usepackage{pgfplots}
 \usetikzlibrary{backgrounds,calc}
 \usepackage{amsmath,tikz}
 \usetikzlibrary{quotes,angles}
\pgfplotsset{compat = newest}
\usetikzlibrary{quotes,angles}

\usetikzlibrary{svg.path}
\usepackage{pgfornament}

\usepackage{xcolor}

\usepackage{subfig}
\usepackage{graphicx}
\def\mset #1[#2]=#3{%
	\expandafter\xdef\csname #1#2\endcsname{#3}
}
\def\mget #1[#2]{%
	\csname #1#2\endcsname
}
\def\minc #1[#2]+=#3{%
	\pgfmathparse{\mget #1[#2]+#3}%
	\mset #1[#2]=\pgfmathresult
}

\date{}

\usepackage{hyperref}
\usepackage [all]{hypcap}
\usepackage{color}

\usepackage{listings}
\definecolor{codegreen}{rgb}{0,0.6,0}
\definecolor{codegray}{rgb}{0.5,0.5,0.5}
\definecolor{codepurple}{rgb}{0.58,0,0.82}
\definecolor{backcolour}{rgb}{0.95,0.95,0.92}

\lstdefinestyle{mystyle}{
    backgroundcolor=\color{backcolour},   
    commentstyle=\color{codegreen},
    keywordstyle=\color{magenta},
    numberstyle=\tiny\color{codegray},
    stringstyle=\color{codepurple},
    basicstyle=\ttfamily\footnotesize,
    breakatwhitespace=false,         
    breaklines=true,                 
    captionpos=b,                    
    keepspaces=true,                 
    numbers=left,                    
    numbersep=5pt,                  
    showspaces=false,                
    showstringspaces=false,
    showtabs=false,                  
    tabsize=2
}

\begin{document}

\title{Alternative Bell's states and Teleportation }
\author{ Juan M. Romero \thanks{jromero@cua.uam.mx }, Emiliano Montoya-Gonz\'alez \thanks{emiliano.montoya.g@cua.uam.mx } and  Oscar Velazquez-Alvarado  \thanks{oscar.velazquez@cua.uam.mx }\\
Departamento de Matemáticas Aplicadas y Sistemas,\\
Universidad Aut\'onoma Metropolitana-Cuajimalpa,\\
M\'exico, D.F 05300, M\'exico }
\date{\today}

\maketitle
\begin{abstract}
Bell's states are among the most useful in quantum computing.  These state are an orthonormal  base of entagled states with  two qubits.
We propose  alternative bases of entangled states. Some of these states depend on a continuous  parameter. We present the quantum circuit and code of these alternative bases. In addition, we study quantum teleportation with these entangled states and present their  quantum circuits and codes  associated .
\end{abstract}

\section{\label{sec:introduction}Introduction}

Recently, quantum computing  has proven to be superior to classical computing. For example,  the  Quantum  Fourier Transform   has $O(n^{2})$  computational complexity \cite{Asaka}, while  the  classical Fast Fourier Transform (FFT)  has computational complexity of $O(n2^{n})$. Then, the  Quantum  Fourier Transform is more efficient than  the classical FFT, a review about quantum computational complexity can be seen in \cite{Wastrous}.   In addition, in quantum cryptography, when  the  Shor's algorithm could be implemented error-free, it would break many classical public-key cryptography schemes  \cite{Lee}. 
 For these reasons, quantum computing development will have a significant impact on simulation to understand various physics systems in high energy \cite{he}, nuclear physics \cite{nuclear}, quantum chemistry \cite{chemistry}, finance  \cite{Herman}, etc.\\

In this respect, in several quantum algorithm,  Bell's states are among the most useful, particularly in quantum  cryptography
 \cite{Lee,bernt,Nishioka,yano}.  These state are an orthonormal  base of entagled states with  two qubits. Furthermore, by using  Bell's state and   another state it is possible to obtain quantum teleportation. 
In this paper we propose  alternative bases of entangled states. Some of these states depend on a continuous parameter. We present the quantum circuit and code of these alternative bases. In addition, we study quantum teleportation with these entangled states and present their  quantum circuits and codes  associated.\\

This paper is organized as follows.  In Sec. \ref{SecBell} we study the general conditions to obtain entangled states. 
In Sec. \ref{SecBell1} we study  general properties of the entangled states.  In Sec. \ref{SecTeleportation} we study quantum teleportation with the alternative bases of  entangled states.   In Sec. \ref{SecExample1} we  present  an example of  alternative base of  entangled states and study teleportation with them, in addition we present   the  quantum circuits and codes  associated. In Secs. \ref{SecExample2}-\ref{SecExample4} we  present other examples of  alternative base of  entangled states.
In Sec. \ref{Con} a summary is given.

\section{Entangled states }
\label{SecBell}

If we have the states
\begin{eqnarray}
\ket{\psi_{1}}&=&\gamma_{1}\ket{0}+\gamma_{2}\ket{1},\nonumber\\
\ket{\psi_{2}}&=&\lambda_{1}\ket{0}+\lambda_{2}\ket{1},\nonumber
\end{eqnarray}
we obtain the state
\begin{eqnarray}
\ket{\psi_{1}}\otimes\ket{\psi_{2}}=\gamma_{1}\lambda_{1}\ket{00}+\gamma_{1}\lambda_{2}\ket{01}
+\gamma_{2}\lambda_{1}\ket{10}+\gamma_{2}\lambda_{2}\ket{11}.\nonumber
\end{eqnarray}
Now, let us define the matrix
\begin{eqnarray}
\begin{pmatrix}
\gamma_{1}\lambda_{1}&\gamma_{1}\lambda_{2}\\
\gamma_{2}\lambda_{1}& \gamma_{2}\lambda_{2}
\end{pmatrix}
\nonumber
\end{eqnarray}
which satisfies 
\begin{eqnarray}
\det \begin{pmatrix}
\gamma_{1}\lambda_{1}&\gamma_{1}\lambda_{2}\\
\gamma_{2}\lambda_{1}& \gamma_{2}\lambda_{2}
\end{pmatrix}  =\gamma_{1}\lambda_{1}\gamma_{2}\lambda_{2}-\gamma_{2}\lambda_{1}\gamma_{1}\lambda_{2}=0
\nonumber
\end{eqnarray}
Then,  if we have the $2$-qubits state
\begin{eqnarray}
\ket{\psi}=c_{1}\ket{00}+c_{2}\ket{01}
+c_{3}\ket{10}+c_{4}\ket{11}\nonumber
\end{eqnarray}
and there are two  $\ket{\psi_{1}}, \ket{\psi_{2}}$ such that 
\begin{eqnarray}
\ket{\psi}=\ket{\psi_{1}}\otimes \ket{\psi_{2}},  \label{state0}
\end{eqnarray}
 the matrix
\begin{eqnarray}
A=\begin{pmatrix}
a_{1}&a_{2}\\
a_{3}& a_{4}
\end{pmatrix}
\nonumber
\end{eqnarray}
satisfies 
\begin{eqnarray}
\det A=0.
\nonumber
\end{eqnarray}
In nother words, if  
\begin{eqnarray}
\det A\not =0.
\label{matriC}
\end{eqnarray}
for all states $\ket{\psi_{1}},\ket{\psi_{2}},$ we have
\begin{eqnarray}
\ket{\psi}\not=\ket{\psi_{1}}\otimes\ket{\psi_{2}}\nonumber
\end{eqnarray}
Then if the equation \ref{matriC}  is satisfied the state (\ref{state0}) represent a mixed state. 
Notice that for each one matrix of  $L(2,\mathbb{C})$ we have  a entangled  state.\\

For example, by using the matrices  
\begin{eqnarray}
B_{1}=\frac{1}{\sqrt{2}}\begin{pmatrix}
1&0\\
0&1
\end{pmatrix},  B_{2}=\frac{1}{\sqrt{2}}\begin{pmatrix}
1&0\\
0&-1
\end{pmatrix},  B_{3}=\frac{1}{\sqrt{2}}\begin{pmatrix}
0&1\\
1&0
\end{pmatrix},  B_{4}=\frac{1}{\sqrt{2}}\begin{pmatrix}
0&1\\
-1&0
\end{pmatrix}, \nonumber
\end{eqnarray}
we obtain the Bell's states
\begin{eqnarray}
\ket{B_{1}}&=&\frac{1}{\sqrt{2}}\left(\ket{00}+\ket{11}\right),\label{sb1} \\
\ket{B_{2}}&=&\frac{1}{\sqrt{2}}\left(\ket{00}-\ket{11}\right),\label{sb2} \\
\ket{B_{3}}&=&\frac{1}{\sqrt{2}}\left(\ket{01}+\ket{10}\right),\label{sb3} \\
\ket{B_{4}}&=&\frac{1}{\sqrt{2}}\left(\ket{01}-\ket{10}\right). \label{sb4} 
\end{eqnarray}

\begin{figure}[hbt!]
\centering
\begin{quantikz}%[color=black,background color=yellow]
\gategroup[wires=2,steps=6,style={rounded corners,fill=blue!20}, background]{}
&\lstick{$|{0}\rangle$} & \gate{H}&\ctrl{1} & \meter{} & \qw
\\
&\lstick{$|{0}\rangle$}  & \qw & \gate{X} \slice{} & \meter{} & \qw
\end{quantikz}
\caption{Quantum circuit for the state \eqref{sb1}. }
    \label{fig:BellCirc1}
\end{figure}
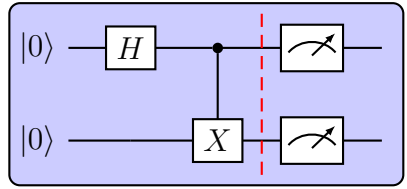

In the figure \ref{fig:BellCirc1} we show  the quantum circuit for the first Bell's state \eqref{sb1} and in the Listing \ref{code:Bell1} we present their code in  Python language using IBM's Qiskit library.

\FloatBarrier

\FloatBarrier 
\begin{lstlisting}[language=Python, caption= Code for the state \eqref{sb1}. , label={code:Bell1}  ]
import qiskit as q

# Create two quantum registers
qr = q.QuantumRegister(2, 'q')

# Create two classic registers
cr = q.ClassicalRegister(2, 'c')

# Create a circuit with the four registers
circuit = q.QuantumCircuit(qr, cr)

# Add the gates to the circuit
circuit.h(qr[0])
circuit.cx(qr[0], qr[1])

# Measure the two qubits
circuit.measure(qr, cr)
\end{lstlisting}

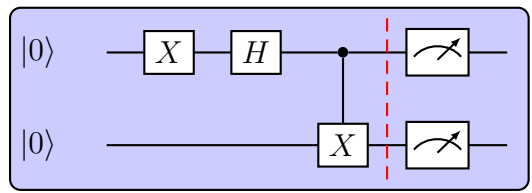
\begin{figure}[hbt!]
\centering
\begin{quantikz}%[color=black,background color=yellow]
\gategroup[wires=2,steps=8,style={rounded corners,fill=blue!20}, background]{}
&\lstick{$|{0}\rangle$} & &\gate{X} & \gate{H}&\ctrl{1} \slice{} & \meter{} & \qw
\\
&\lstick{$|{0}\rangle$}  & & \qw & \qw & \gate{X} & \meter{} & \qw
\end{quantikz}
\caption{Quantum circuit for the state   \eqref{sb2}.}
    \label{fig:BellCirc2}
\end{figure}

In the figure \ref{fig:BellCirc2} we show  the quantum circuit for the second Bell´s states and in the Listing \ref{code:Bell2} we present their code in  Python language using IBM's Qiskit library.
\FloatBarrier

% Segundo Bell
\FloatBarrier 
\begin{lstlisting}[language=Python, caption=Code for the state \eqref{sb2}., label={code:Bell2}  ]
import qiskit as q

# Create two quantum registers
qr = q.QuantumRegister(2, 'q')

# Create two classic registers
cr = q.ClassicalRegister(2, 'c')

# Create a circuit with the four registers
circuit = q.QuantumCircuit(qr, cr)

# Add the gates to the circuit
circuit.x(qr[0])
circuit.h(qr[0])
circuit.cx(qr[0], qr[1])

# Measure the two qubits
circuit.measure(qr, cr)
\end{lstlisting}

\begin{figure}[hbt!]
\centering
\begin{quantikz}%[color=black,background color=yellow]
\gategroup[wires=2,steps=7,style={rounded corners,fill=blue!20}, background]{}
&\lstick{$|{0}\rangle$} & & \gate{H}&\ctrl{1} \slice{} & \meter{} & \qw
\\
&\lstick{$|{0}\rangle$}  & & \gate{X} & \gate{X} & \meter{} & \qw
\end{quantikz}
\caption{Quantum circuit for the state   \eqref{sb3}.}
    \label{fig:BellCirc3} 
\end{figure}
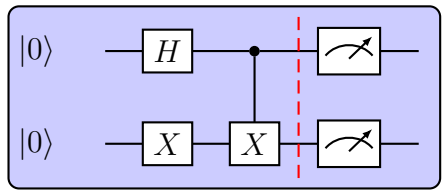

In the figure \ref{fig:BellCirc3} we show  the quantum circuit for the third Bel's states and in the Listing \ref{code:Bell3} we present their code in  Python language using IBM's Qiskit library.

\FloatBarrier 

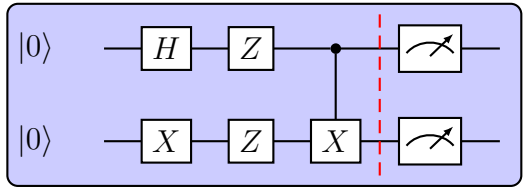
\begin{figure}[hbt!]
\centering
\begin{quantikz}%[color=black,background color=yellow]
\gategroup[wires=2,steps=8,style={rounded corners,fill=blue!20}, background]{}
&\lstick{$|{0}\rangle$} & & \gate{H}& \gate{Z} & \ctrl{1} \slice{} & \meter{} & \qw
\\
&\lstick{$|{0}\rangle$}  & & \gate{X} & \gate{Z} & \gate{X} & \meter{} & \qw
\end{quantikz}
\caption{Quantum circuit for the state   \eqref{sb4}. }
    \label{fig:BellCirc4}
\end{figure}

\FloatBarrier 
\begin{lstlisting}[language=Python, caption=Code for the state \eqref{sb3}., label={code:Bell3}  ]
import qiskit as q

# Create two quantum registers
qr = q.QuantumRegister(2, 'q')

# Create two classic registers
cr = q.ClassicalRegister(2, 'c')

# Create a circuit with the four registers
circuit = q.QuantumCircuit(qr, cr)

# Add the gates to the circuit
circuit.h(qr[0])
circuit.x(qr[1])
circuit.cx(qr[0], qr[1])

# Measure the two qubits
circuit.measure(qr, cr)
\end{lstlisting}
In the figure \ref{fig:BellCirc4} we show  the quantum circuit for the fourth Bell's states and in the Listing \ref{code:Bell4} we present their code in  Python language using IBM's Qiskit library.

\FloatBarrier

\begin{lstlisting}[language=Python, caption=Code for the state \eqref{sb4}., label={code:Bell4}  ]
import qiskit as q

# Create two quantum registers
qr = q.QuantumRegister(2, 'q')

# Create two classic registers
cr = q.ClassicalRegister(2, 'c')

# Create a circuit with the four registers
circuit = q.QuantumCircuit(qr, cr)

# Add the gates to the circuit
circuit.h(qr[0])
circuit.z(qr[0])
circuit.x(qr[1])
circuit.z(qr[1])
circuit.cx(qr[0], qr[1])

# Measure the two qubits
circuit.measure(qr, cr)
\end{lstlisting}

 \section{General case}
\label{SecBell1}

In general, if we have the following matrices 
\begin{eqnarray}
A_{0}=\begin{pmatrix}
a_{00} & a_{01}\\
a_{02}& a_{03}
\end{pmatrix},
 A_{1}=\begin{pmatrix}
a_{10} & a_{11}\\
a_{12}& a_{13}
\end{pmatrix},
A_{2}=\begin{pmatrix}
a_{20} & a_{21}\\
a_{22}& a_{23}
\end{pmatrix},
A_{3}=\begin{pmatrix}
a_{30} & a_{31}\\
a_{32}& a_{33}
\end{pmatrix}, \label{GM}
\end{eqnarray}
where 
\begin{eqnarray}
\det A_{i}=
a_{i0}a_{i3}- a_{i1}a_{i2}\not =0,\quad a_{ij}\in \mathbb{C},ij=0,1,2,3
\end{eqnarray}
 we can propose the entangled states 
\begin{eqnarray}
\ket{V_{0}}&=&a_{00}\ket{00}+a_{01}\ket{01}  +a_{02}\ket{10}  +a_{03}\ket{11},      \nonumber \\
\ket{V_{1}}&=&a_{10}\ket{00}+a_{11}\ket{01}  +a_{12}\ket{10}  +a_{13}\ket{11} ,     \nonumber \\
\ket{V_{2}}&=&a_{20}\ket{00}+a_{21}\ket{01}  +a_{22}\ket{10}  +a_{23}\ket{11},      \nonumber \\
\ket{V_{3}}&=&a_{30}\ket{00}+a_{31}\ket{01}  +a_{32}\ket{10}  +a_{33}\ket{11}.      \nonumber 
\end{eqnarray}
Notices that to obtain  a base of orthonormal enangled  states $\ket{V_{j}}$ the equation 
\begin{eqnarray}
\braket{V_{i}|V_{j}}&=&a_{i0}^{*}a_{j0}+a_{i1}^{*}a_{j1}+a_{i2}^{*}a_{j2}+a_{i3}^{*}a_{j3}=\delta_{ij},\nonumber
\end{eqnarray}
must be satisfied. This equation which  can be written as 
\begin{eqnarray}
Tr\left( A^{\dagger}_{i}A_{j} \right)=\braket{V_{i}|V_{j}}=\delta_{ij}. \nonumber
\end{eqnarray}

In addition, notice that by  using the matrix
\begin{eqnarray}
T=
\begin{pmatrix}
a_{00}&a_{01}&a_{02}&a_{03}\\
a_{10}&a_{11}&a_{12}&a_{13}\\
a_{20}&a_{21}&a_{22}&a_{23}\\
a_{30}&a_{31}&a_{32}&a_{3}\\
\end{pmatrix}
\end{eqnarray}
 the states $\ket{V_{i}}$  can be written as
\begin{eqnarray}
\begin{pmatrix}
 \ket{V_{0}}\\
 \ket{V_{1}}\\
 \ket{V_{2}}\\
 \ket{V_{3}}
\end{pmatrix}=
\begin{pmatrix}
a_{00}&a_{01}&a_{02}&a_{03}\\
a_{10}&a_{11}&a_{12}&a_{13}\\
a_{20}&a_{21}&a_{22}&a_{23}\\
a_{30}&a_{31}&a_{32}&a_{3}\\
\end{pmatrix}
\begin{pmatrix}
\ket{00}\\
\ket{01}\\
\ket{10}\\
\ket{11}
\end{pmatrix},
\end{eqnarray}
namely
\begin{eqnarray}
\ket{V_{i}}=T_{ij}\ket{C_{j}}, \label{NBell}
\end{eqnarray}
where
\begin{eqnarray}
\ket{C_{0}}&=&\ket{00},\\
\ket{C_{1}}&=&\ket{01},\\
\ket{C_{2}}&=&\ket{10},\\
\ket{C_{3}}&=&\ket{11}.
\end{eqnarray}
By using the properties of the matrices \eqref{GM}  it can be shown that $T$ satisfies 
\begin{eqnarray}
T^{-1}=T^{\dagger}=T^{*T},\nonumber
\end{eqnarray}
namely   $T$ is an unitary matrix.\\

 In addition, the equation \eqref{NBell}   implies 
\begin{eqnarray}
\ket{C_j}=T^{-1}_{jk}\ket{V_{k}}
\end{eqnarray}
which can be written as
\begin{eqnarray}
\ket{C_j}=a_{kj}^{*}\ket{V_{k}}. \label{cb1}
\end{eqnarray}

\section{Quantum teleportation}
\label{SecTeleportation}

The basic teleportation algorithm consider three qubits. The first one is the target qubit, the second one is the Alice qubit and the third one is the Bob qubit.\\

In order to study this subject, notice that  by using the state
\begin{eqnarray}
\ket{\psi}=\gamma_{1}\ket{0}+\gamma_{2}\ket{1},
\end{eqnarray}
we have 
\begin{eqnarray}
\ket{\phi_{i}}&=&\ket{\psi}\otimes \ket{V_{i}}\nonumber\\
&=&\left(\gamma_{1}\ket{0}+\gamma_{2}\ket{1}\right)\otimes\left(a_{ij}\ket{C_j}\right)\nonumber\\
&=&\gamma_{1}\ket{0}\otimes a_{ij}\ket{C_j}+\gamma_{2}\ket{1}\otimes a_{ij}\ket{C_j}\nonumber\\
&=&\gamma_{1}\left(a_{i0}\ket{0}\ket{00}+a_{i1}\ket{0}\ket{01}+a_{i2}\ket{0}\ket{10}+a_{i3}\ket{0}\ket{11}\right)+ \nonumber\\
& &+\gamma_{2}\left(a_{i0}\ket{1}\ket{00}+a_{i1}\ket{1}\ket{01}+a_{i2}\ket{1}\ket{10}+a_{i3}\ket{1}\ket{11}\right) \nonumber\\
&=&\ket{00}\left(\gamma_{1}a_{i0}\ket{0}+\gamma_{1}a_{i1}\ket{1}\right)+\ket{01}\left( \gamma_{1}a_{i2}\ket{0}+\gamma_{1}a_{i3}\ket{1}\right)+ \nonumber\\
& &+\ket{10}\left(\gamma_{2}a_{i0}\ket{0}+\gamma_{2}a_{i1}\ket{1}\right)+\ket{11} \left(\gamma_{2}a_{i2}\ket{0}+\gamma_{2}a_{i3}\ket{1}\right) \nonumber
\end{eqnarray}
which can be written as
\begin{eqnarray}
\ket{\phi_{i}}&=&\ket{C_0}\left(\gamma_{1}a_{i0}\ket{0}+\gamma_{1}a_{i1}\ket{1}\right)+\ket{C_1}\left( \gamma_{1}a_{i2}\ket{0}+\gamma_{1}a_{i3}\ket{1}\right)+ \nonumber\\
& &+\ket{C_2}\left(\gamma_{2}a_{i0}\ket{0}+\gamma_{2}a_{i1}\ket{1}\right)+\ket{C_3} \left(\gamma_{2}a_{i2}\ket{0}+\gamma_{2}a_{i3}\ket{1}\right). \nonumber
\end{eqnarray}
Furthermore by using the equation \eqref{cb1}  we obtain
\begin{eqnarray}
\ket{\phi_{i}}&=&\ket{V_{k}}\Big(\gamma^{\prime}_{1ki}\ket{0}+\gamma^{\prime}_{2ki}\ket{1} \Bigg),\nonumber
\end{eqnarray}
where
\begin{eqnarray}
\begin{pmatrix}
\gamma_{1ki}^{\prime}\\
\gamma_{2ki}^{\prime}
\end{pmatrix}
=\left(A_{k}^{*}A_{i}\right)^{T}
\begin{pmatrix}
\gamma_{1}\\
\gamma_{2}
\end{pmatrix}. \nonumber
\end{eqnarray}
Then 
\begin{eqnarray}
\ket{\phi_{i}}&=&\ket{\psi}\ket{V_{i}}\nonumber\\
&=&\ket{V_{0}}\left(\gamma_{10i}^{\prime}\ket{0}+\gamma^{\prime}_{20i}\ket{1}\right)+\ket{V_{1}}\left(\gamma^{\prime}_{11i}\ket{0}+\gamma^{\prime}_{21i}\ket{1}\right)+\nonumber\\
& &
+\ket{V_{2}}\left(\gamma^{\prime}_{12i}\ket{0}+\gamma^{\prime}_{22i}\ket{1}\right)+\ket{V_{3}}\left(\gamma^{\prime}_{13i}\ket{0}+\gamma^{\prime}_{23i}\ket{1}\right)\nonumber
\end{eqnarray}
where
\begin{eqnarray}
\begin{pmatrix}
\gamma_{10i}^{\prime}\\
\gamma_{20i}^{\prime}
\end{pmatrix}
&=&A^{T}_{i}A_{0}^{*T}
\begin{pmatrix}
\gamma_{1}\\
\gamma_{2}
\end{pmatrix},\nonumber\\
\begin{pmatrix}
\gamma_{11i}^{\prime}\\
\gamma_{21i}^{\prime}
\end{pmatrix}
&=&A^{T}_{i}A_{1}^{*T}
\begin{pmatrix}
\gamma_{1}\\
\gamma_{2}
\end{pmatrix},\nonumber\\
\begin{pmatrix}
\gamma_{12i}^{\prime}\\
\gamma_{22i}^{\prime}
\end{pmatrix}
&=&A^{T}_{i}A_{3}^{*T}
\begin{pmatrix}
\gamma_{1}\\
\gamma_{2}
\end{pmatrix},\nonumber\\
\begin{pmatrix}
\gamma_{13i}^{\prime}\\
\gamma_{23i}^{\prime}
\end{pmatrix}
&=&A^{T}_{i}A_{4}^{*T}
\begin{pmatrix}
\gamma_{1}\\
\gamma_{2}
\end{pmatrix}.\nonumber
\end{eqnarray}

Now, if a observer $O_{1}$ measure the state
$\ket{V_{k}}$
the observer $O_{2}$ have to applied the gate 
\begin{eqnarray}
\left(\left(A_{k}^{*}A_{i}\right)^{T}\right)^{-1}.
\end{eqnarray}

\section{Example 1}
\label{SecExample1}

For example, by using the orthonormal and invertible  matrices  
\begin{eqnarray}
A_{1}&=&\frac{1}{\sqrt{2}}\begin{pmatrix}
e^{i\theta} &0 \\
0&1
\end{pmatrix}, \label{mExp1}  \\
A_{2}&=&\frac{1}{\sqrt{2}}\begin{pmatrix}
1 &0\\
0 &-e^{-i\theta}
\end{pmatrix},\label{mExp2} \\
  A_{3}&=&\frac{1}{\sqrt{2}}\begin{pmatrix}
0&1\\
-1&0
\end{pmatrix}, \label{mExp3}  \\
A_{4}&=&\frac{1}{\sqrt{2}}\begin{pmatrix}
0&1\\
1&0
\end{pmatrix}, \label{mExp4}
\end{eqnarray}
the following orthonormal  basis  of entagled  states 
\begin{eqnarray}
\ket{V_{1}}&=&\frac{1}{\sqrt{2}}\left(e^{i \theta} \ket{00}+ \ket{11}\right),\label{sbExpI1} \\
\ket{V_{2}}&=&\frac{1}{\sqrt{2}}\left( \ket{00}-e^{-i\theta}  \ket{11}\right),\label{sbExpI2}  \\
\ket{V_{3}}&=&\frac{1}{\sqrt{2}}\left(\ket{01}-\ket{10}\right),\label{sbExpI3} \\
\ket{V_{4}}&=&\frac{1}{\sqrt{2}}\left(\ket{01}+\ket{10}\right). \label{sbExpI4} 
\end{eqnarray}
can be constructed.\\

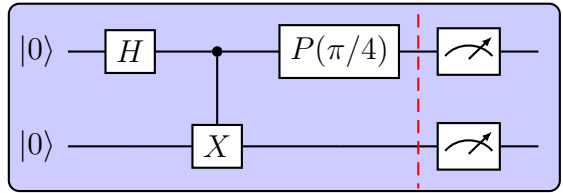
\begin{figure}[hbt!]
\centering
\begin{quantikz}%[color=black,background color=yellow]
\gategroup[wires=2,steps=7,style={rounded corners,fill=blue!20}, background]{}
&\lstick{$|{0}\rangle$} & \gate{H}& \ctrl{1} & \gate{P(\pi /4)}\slice{} & \meter{} & \qw
\\
&\lstick{$|{0}\rangle$}  & \qw & \gate{X} & \qw & \meter{} & \qw
\end{quantikz}
\caption{Quantum circuit for the state \eqref{sbExpI1}. }
    \label{fig:BellExpI1}
\end{figure}

In the figure \ref{fig:BellExpI1} we show  the quantum circuit for the  states \eqref{sbExpI1} and in the Listing \ref{code:BellExpI1} we present their code in  Python language using IBM's Qiskit library.

\FloatBarrier 

\begin{lstlisting}[language=Python, caption=Code  for the state \eqref{sbExpI1}. , label={code:BellExpI1}  ]
import numpy as np
import matplotlib.pyplot as plt
from qiskit import QuantumCircuit, QuantumRegister
import qiskit.quantum_info as qi

qreg_q = QuantumRegister(2, 'q')
creg_c = ClassicalRegister(4, 'c')
circuit = QuantumCircuit(qreg_q, creg_c)

circuit.h(qreg_q[0])
circuit.cx(qreg_q[0], qreg_q[1])
circuit.p(np.pi/4, qreg_q[0])
circuit.swap(qreg_q[0], qreg_q[1])
\end{lstlisting}

\begin{figure}[hbt!]
\centering
\begin{quantikz}%[color=black,background color=yellow]
\gategroup[wires=2,steps=9,style={rounded corners,fill=blue!20}, background]{}
&\lstick{$|{0}\rangle$} & \qw & \gate{H} & \qw &\ctrl{1} & \qw \slice{} & \meter{} & \qw
\\
&\lstick{$|{0}\rangle$}  & \qw & \qw & \gate{Sdg} & \gate{X} & \gate{P(-\pi /4)} & \meter{} & \qw
\end{quantikz}
\caption{Quantum circuit for the state \eqref{sbExpI2}.  }
    \label{fig:BellExpI2}
\end{figure}
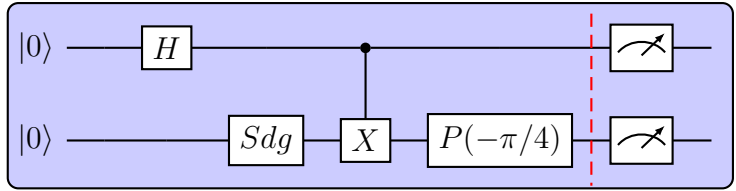

In the figure \ref{fig:BellExpI2} we show  the quantum circuit for the  state \eqref{sbExpI2} and in the Listing \ref{code:BellExpI2} we present their code in  Python language using IBM's Qiskit library.

\FloatBarrier 

\begin{lstlisting}[language=Python, caption=Code  for the state \eqref{sbExpI2}. , label={code:BellExpI2}  ]
import numpy as np
import matplotlib.pyplot as plt
from qiskit import QuantumCircuit, QuantumRegister
import qiskit.quantum_info as qi

qreg_q = QuantumRegister(2, 'q')
creg_c = ClassicalRegister(4, 'c')
circuit = QuantumCircuit(qreg_q, creg_c)

circuit.h(qreg_q[0])
circuit.barrier(qreg_q[0], qreg_q[1])
circuit.sdg(qreg_q[1])
circuit.cx(qreg_q[0], qreg_q[1])
circuit.p(-pi / 4, qreg_q[1])
circuit.barrier(qreg_q[0], qreg_q[1])
circuit.swap(qreg_q[0], qreg_q[1])
\end{lstlisting}

Notice que the states  \eqref{sbExpI3}  and  \eqref{sbExpI4}  are the states   \eqref{sb4} and \eqref{sb3}, the quantum circuits and codes for these states are on the Section \ref{SecBell}.\\

In addition, by using the matrices \eqref{mExp1}-\eqref{mExp4}, we obtain the matrix

\begin{eqnarray}
T=\frac{1}{\sqrt{2}}\begin{pmatrix}
e^{i \theta} &0&0 &1\\
1 &0&0&-e^{-i\theta}\\
0&1&-1&0\\
0&1&1&0
\end{pmatrix}
\end{eqnarray}
and their inverse 
\begin{eqnarray}
T^{-1}=T^{\dagger}=T^{*T}=\frac{1}{\sqrt{2}}\begin{pmatrix}
e^{-i \theta} &1&0 &0\\
0 &0&1&1\\
0&0&-1&1\\
1&-e^{i\theta}&0&0
\end{pmatrix}
\end{eqnarray}
Notice that the states \eqref{sbExpI1}-\eqref{sbExpI4} and the matrix $T^{-1}$ imply the equations 
\begin{eqnarray}
\ket{00}&=&\frac{1}{\sqrt{2}}
\left(e^{-i\theta} \ket{V_{1}}+ \ket{V_{2}}\right),\label{InvExpI1} \\
\ket{01}&=&\frac{1}{\sqrt{2}}\left(
\ket{V_{3}}+\ket{V_{4}}\right),\label{InvExpI2} \\
\ket{10}&=&\frac{1}{\sqrt{2}}\left(-\ket{V_{3}}+\ket{V_{4}}\right),\label{InvExpI3}\\
\ket{11}&=&\frac{1}{\sqrt{2}}\left( \ket{V_{1}}
-e^{i \theta} \ket{V_{2}}\right).\label{InvExpI4}
\end{eqnarray}

Furthermore, by using the matrices  \eqref{mExp1}-\eqref{mExp4}, we have
\begin{eqnarray}
A_{1}^{T}A_{1}^{*T}&=&\frac{1}{2}\begin{pmatrix}
    1&0\\
    0& 1
\end{pmatrix},\nonumber\\
A_{1}^{T}A_{2}^{*T}&=&\frac{1}{2}\begin{pmatrix}
    e^{i\theta}&0 \\
    0& - e^{i\theta}
\end{pmatrix}\nonumber\\
A_{1}^{T}A_{3}^{*T}&=&\frac{1}{2}\begin{pmatrix}
    0&-e^{i\theta} \\
    1& 0
\end{pmatrix},\nonumber\\
A_{1}^{T}A_{4}^{*T}&=&\frac{1}{2}\begin{pmatrix}
    0&e^{i\theta} \\
    1& 0
\end{pmatrix}\nonumber\\
A_{2}^{T}A_{1}^{*T}&=&\frac{1}{2}\begin{pmatrix}
    e^{-i\theta}&0\\
    0&  -e^{-i\theta}
\end{pmatrix}\nonumber\\
A_{2}^{T}A_{2}^{*T}&=&\frac{1}{2}\begin{pmatrix}
    1&0 \\
    0& 1
\end{pmatrix}\nonumber\\
A_{2}^{T}A_{3}^{*T}&=&\frac{1}{2}\begin{pmatrix}
    0&-1 \\
    -e^{-i\theta}& 0
\end{pmatrix}\nonumber\\
A_{2}^{T}A_{4}^{*T}&=&\frac{1}{2}\begin{pmatrix}
    0&1 \\
    -e^{-i\theta}& 0
\end{pmatrix}\nonumber\\
A_{3}^{T}A_{1}^{*T}&=&\frac{1}{2}\begin{pmatrix}
    0&-1 \\
    e^{-i\theta}& 0
\end{pmatrix}\nonumber\\
A_{3}^{T}A_{2}^{*T}&=&\frac{1}{2}\begin{pmatrix}
    0&e^{i\theta} \\
    1& 0
\end{pmatrix}\nonumber\\
A_{3}^{T}A_{3}^{*T}&=&\frac{1}{2}\begin{pmatrix}
    -1&0 \\
    0& -1
\end{pmatrix}\nonumber\\
A_{3}^{T}A_{4}^{*T}&=&\frac{1}{2}\begin{pmatrix}
    -1&0\\
    0& 1
\end{pmatrix}\nonumber\\
A_{4}^{T}A_{1}^{*T}&=&\frac{1}{2}\begin{pmatrix}
    0&1\\
    e^{-i\theta}& 0
\end{pmatrix}\nonumber\\
A_{4}^{T}A_{2}^{*T}&=&\frac{1}{2}\begin{pmatrix}
    0&-e^{i\theta}\\
    1& 0
\end{pmatrix}\nonumber\\
A_{4}^{T}A_{3}^{*T}&=&\frac{1}{2}\begin{pmatrix}
    1&0\\
    0& -1
\end{pmatrix}\nonumber\\
A_{4}^{T}A_{4}^{*T}&=&\frac{1}{2}\begin{pmatrix}
    1&0\\
    0& 1
\end{pmatrix}.  \nonumber
\end{eqnarray}

Now, by using these last matrices and  the equations  \eqref{InvExpI1}-\eqref{InvExpI4} we obtain the following states 
\begin{eqnarray}
\ket{\phi_{1}}&=&\ket{\psi}\otimes\ket{V_{1}} \nonumber\\
&=&\ket{V_{1}}\frac{1}{2}\left(  \alpha_{1}\ket{0}+ \alpha_{2}\ket{1} \right)+\nonumber\\
& &+
\ket{V_{2}}\frac{1}{2}\left(  e^{i\theta}\alpha_{1}\ket{0}- e^{i\theta}\alpha_{2}\ket{1} \right)+\nonumber\\
& &+\ket{V_{3}}\frac{1}{2}\left( - e^{i\theta}\alpha_{2}\ket{0}+\alpha_{1}\ket{1} \right)+\nonumber\\
& &+\ket{V_{4}}\frac{1}{2}\left( e^{i\theta}\alpha_{2}\ket{0}+ \alpha_{1}\ket{1} \right)\label{STEpI1}\\
\ket{\phi_{2}}&=&\ket{\psi}\otimes\ket{V_{2}} \nonumber\\
&=&\ket{V_{1}}\frac{1}{2}\left(  e^{-i\theta}\alpha_{1}\ket{0}- e^{-i\theta}\alpha_{2}\ket{1} \right)+\nonumber\\\
& &+
\ket{V_{2}}\frac{1}{2}\left(  \alpha_{1}\ket{0}+\alpha_{2}\ket{1} \right)+\nonumber\\
& &+\ket{V_{3}}\frac{1}{2}\left(  -\alpha_{2}\ket{0}- e^{-i\theta}\alpha_{1}\ket{1} \right)+\nonumber\\
& &+\ket{V_{4}}\frac{1}{2}\left(  \alpha_{2}\ket{0}- e^{-i\theta}\alpha_{1}\ket{1} \right)\label{STEpI2}\\
\ket{\phi_{3}}&=&\ket{\psi}\otimes\ket{V_{3}} \nonumber\\
&=&\ket{V_{1}}\frac{1}{2}\left(  -\alpha_{2}\ket{0}+ e^{-i\theta}\alpha_{1}\ket{1} \right)+\nonumber\\
& &+
\ket{V_{2}}\frac{1}{2}\left(  e^{i\theta}\alpha_{2}\ket{0}+\alpha_{1}\ket{1} \right)+\nonumber\\
& &+\ket{V_{3}}\frac{(-1)}{2}\left(  \alpha_{1}\ket{0}+ \alpha_{2}\ket{1} \right)+\nonumber\\
& &+\ket{V_{4}}\frac{1}{2}\left(  -\alpha_{1}\ket{0} +\alpha_{2}\ket{1} \right)\label{STEpI3}\\
\ket{\phi_{4}}&=&\ket{\psi}\otimes\ket{V_{4}} \nonumber\\
&=&\ket{V_{1}}\frac{1}{2}\left(  \alpha_{2}\ket{0}+ e^{-i\theta}\alpha_{1}\ket{1} \right)+\nonumber\\
& &+
\ket{V_{2}}\frac{1}{2}\left(  -e^{i\theta}\alpha_{2}\ket{0}+ \alpha_{1}\ket{1} \right)+\nonumber\\
& &+\ket{V_{3}}\frac{1}{2}\left(  \alpha_{1}\ket{0}-\alpha_{2}\ket{1} \right)+\ket{V_{4}}\frac{1}{2}\left(  \alpha_{1}\ket{0}+
\alpha_{2}\ket{1} \right).\label{STEpI4}
\end{eqnarray}

\begin{figure}[hbt!]
\centering
\begin{quantikz}%[color=black,background color=yellow]
\gategroup[wires=3,steps=13,style={rounded corners,fill=blue!20}, background]{}
&\lstick{$|{0}\rangle$} & \qw & \qw
 \slice{} & \gate{H} \slice{} & \ctrl{1} & \gate{H} & \ctrl{1} & \ctrl{1} \slice{} & \meter{} \slice{} & \qw & \ctrl{2} & \qw
\\
&\lstick{$|{0}\rangle$}  & \gate{H} & \ctrl{1} & \qw & \gate{X} & \qw & \gate{P(\pi / 4)} & \gate{P(-\pi / 4)} & \meter{} & \ctrl{1} & \qw & \qw
\\
&\lstick{$|{0}\rangle$}  & \qw & \gate{X} & \qw & \qw & \qw & \qw & \qw\qw & \qw & \gate{X} & \gate{Z} & \qw
\end{quantikz}
\caption{Quantum circuit  for  teleportation with the states  \eqref{STEpI1}- \eqref{STEpI4}. }
    \label{fig:BellImg}
\end{figure}
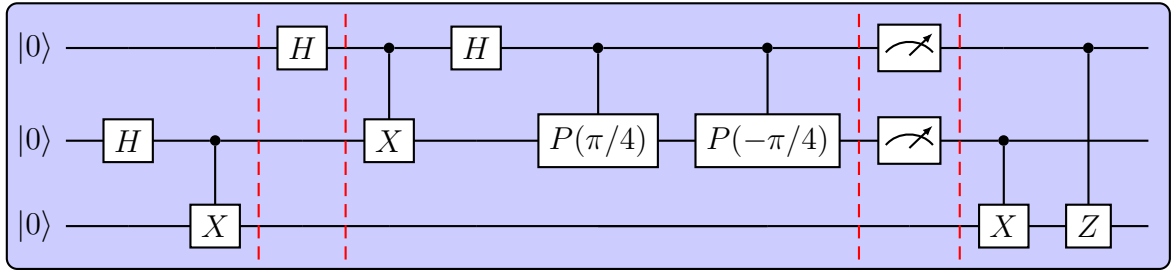

In the Figure \ref{fig:BellImg} we present the quantum circuit  for  teleportation with the states  \eqref{STEpI1}- \eqref{STEpI4} and in the  Listing \ref{code:TExpI} we present the asociated code in  Python language using IBM's Qiskit library.

\FloatBarrier

\begin{lstlisting}[language=Python, caption=Code  for  teleportation with the states  \eqref{STEpI1}- \eqref{STEpI4}. , label={code:TExpI}  ]
from qiskit import QuantumRegister, ClassicalRegister, 
from numpy import pi
import math

qreg_q = QuantumRegister(3, 'q')
creg_c = ClassicalRegister(2, 'c')
qc = QuantumCircuit(qreg_q, creg_c)

qc.h(1)
qc.cx(1, 2) 
qc.x(0);
qc.cx(0, 1)  
qc.h(0)  
theta = pi/4 
qc.cp(theta, 0, 1) 
qc.cp(-theta, 1, 0) 

qc.measure([0, 1], [0, 1])

qc.x(2).c_if(qc.clbits[0], 1)  
qc.z(2).c_if(qc.clbits[1], 1) 
\end{lstlisting}
\FloatBarrier

The algorithm given in Listing \ref{code:TExpI} works as follows:
\begin{enumerate}
\item Entangle the Alice and Bob qubits (is the same entanglement we see in basic Bell states).
\item Prepare the state we want to teleport.
\item In the original algorithm, we apply the Bell Measurement, which means we will collapse the target qubit into one of the Bell states. In the generalized algorithm, we must adapt this circuit to the basis we are working with. 
\item Apply the classic measurement of the target and Alice qubits.
\item Apply the corrections using Controlled NOT and Controlled Z gates. This parts works for every basis.
\end{enumerate}

\section{Example 2}
\label{SecExample2}

For example, using the matrices  
\begin{eqnarray}
A_{1}&=&\frac{1}{\sqrt{2}}\begin{pmatrix}
\cos \theta &-\sin \theta \\
\sin \theta &\cos\theta  
\end{pmatrix},  \nonumber\\
A_{2}&=&\frac{1}{\sqrt{2}}\begin{pmatrix}
-\sin \theta  &-\cos \theta \\
\cos\theta &-\sin\theta
\end{pmatrix},   \nonumber\\  
A_{3}&=&\frac{1}{\sqrt{2}}\begin{pmatrix}
1&0\\
0&-1
\end{pmatrix}, \nonumber\\ 
 A_{4}&=&\frac{1}{\sqrt{2}}\begin{pmatrix}
0&1\\
1&0
\end{pmatrix}, \nonumber
\end{eqnarray}
the Bell's base can be constructed 
\begin{eqnarray}
\ket{V_{1}}&=&\frac{1}{\sqrt{2}}\left(\cos \theta \ket{00}-\sin \theta \ket{01}+\sin \theta \ket{10}+\cos \theta\ket{11}\right),\label{sbRot1} \\
\ket{V_{2}}&=&\frac{1}{\sqrt{2}}\left(-\sin \theta \ket{00}-\cos \theta \ket{01}+\cos \theta\ket{10}-\sin \theta \ket{11}\right),\label{sbRot2} \\
\ket{V_{3}}&=&\frac{1}{\sqrt{2}}\left(\ket{00}-\ket{11}\right),\label{sbRot3} \\
\ket{V_{4}}&=&\frac{1}{\sqrt{2}}\left(\ket{01}+\ket{10}\right). \label{sbRot4} 
\end{eqnarray}

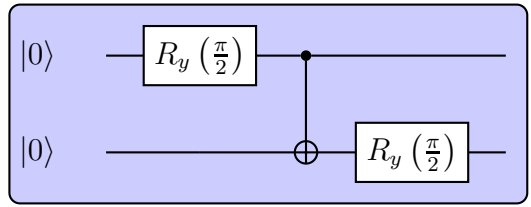
\begin{figure}[hbt!]
\centering
\begin{quantikz}
\gategroup[wires=2,steps=7,style={rounded corners,fill=blue!20}, background]{} 
&\lstick{$|{0}\rangle$} & & \gate{R_y\left(\frac{\pi}{2}\right)} & \ctrl{1} & \qw & \qw \\
&\lstick{$|{0}\rangle$} & & \qw & \targ{} & \gate{R_y\left(\frac{\pi}{2}\right)} & \qw
\end{quantikz}
\caption{Quantum circuit for the state \eqref{sbRot1}.}
\label{fig:BellRot1}
\end{figure}

In the figure \ref{fig:BellRot1} we show  the quantum circuit for the  states \eqref{sbRot1} and in the Listing \ref{code:BellRot1} we present their code in  Python language using IBM's Qiskit library.

\FloatBarrier 
\begin{lstlisting}[language=Python, caption=Code for the state \eqref{sbRot1} , label={code:BellRot1}  ]
from qiskit import QuantumCircuit, Aer, execute
from math import pi, cos, sin

def create_circuit(theta):
    qc = QuantumCircuit(2)
    qc.ry(2 * theta, 0)
    qc.cx(0, 1)
    qc.ry(2 * theta, 1)
    return qc

theta = pi/4
\end{lstlisting}

\begin{figure}[hbt!]
\centering
\begin{quantikz}
\gategroup[wires=2,steps=7,style={rounded corners,fill=blue!20}, background]{} 
&\lstick{$|{0}\rangle$} & & \gate{R_y\left(\frac{\pi}{2}\right)} & \ctrl{1} & \gate{Z} & \qw \\
&\lstick{$|{0}\rangle$} & & & \gate{R_y\left(\frac{\pi}{2}\right)} & \gate{Z} & \qw
\end{quantikz}
\caption{Quantum Circuit for the state \eqref{sbRot2}. }
\label{fig:BellRot2}
\end{figure}
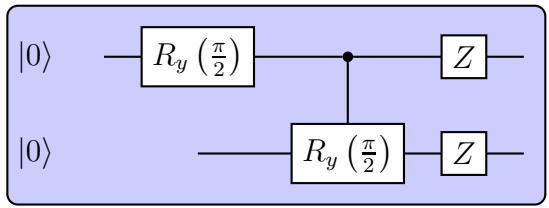

In the figure \ref{fig:BellRot2} we show  the quantum circuit for the  states \eqref{sbRot2}  and in the Listing \ref{code:BellRot2} we present their code in  Python language using IBM's Qiskit library.

\begin{lstlisting}[language=Python, caption=Code for the state \eqref{sbRot2}.  , label={code:BellRot2}  ]
from qiskit import QuantumCircuit, Aer, execute
from math import pi, cos, sin
import numpy as np

def create_circuit(theta):
    qc = QuantumCircuit(2)
    qc.ry(2 * theta, 0)
    qc.cx(0, 1)
    qc.ry(2 * theta, 1)
    qc.z(0)
    qc.z(1)
    return qc

theta = pi / 4
\end{lstlisting}

Notice que the states  \eqref{sbRot3} and  \eqref{sbRot4} are the states   \eqref{sb2} and \eqref{sb3}, the quantum circuits and codes are on the Section \ref{SecBell}.\\

In addition, in this case we have the matrix
\begin{eqnarray}
T=\frac{1}{\sqrt{2}}\begin{pmatrix}
\cos \theta &-\sin\theta&\sin \theta &\cos \theta\\
-\sin \theta &-\cos \theta &\cos\theta&-\sin \theta\\
1&0&0&-1\\
0&1&1&0
\end{pmatrix}
\end{eqnarray}
and 
\begin{eqnarray}
T^{-1}=T^{\dagger}=T^{T}=\frac{1}{\sqrt{2}}\begin{pmatrix}
\cos \theta &-\sin \theta & 1& 0\\
-\sin \theta &-\cos \theta &0&1\\
\sin \theta &\cos \theta &0&1\\
\cos \theta &-\sin \theta &-1&0
\end{pmatrix}
\end{eqnarray}
Then 
\begin{eqnarray}
\ket{00}&=&\frac{1}{\sqrt{2}}\left(\cos \theta \ket{V_{1}}-\sin \theta \ket{V_{2}}+ \ket{V_{3}}\right),\label{InvRot1} \\
\ket{01}&=&\frac{1}{\sqrt{2}}\left(-\sin \theta \ket{V_{1}}-\cos \theta\ket{V_{2}}+ \ket{V_{4}}\right), \label{InvRot2} \\
\ket{10}&=&\frac{1}{\sqrt{2}}\left(\sin \theta \ket{V_{1}}+\cos \theta\ket{V_{2}}+\ket{V_{4}}\right),\label{InvRot3} \\
\ket{11}&=&\frac{1}{\sqrt{2}}\left(\cos \theta \ket{V_{1}}-\sin \theta \ket{V_{2}}-\ket{V_{3}}\right). \label{InvRot4} 
\end{eqnarray}

In addition we have
\begin{eqnarray}
A_{1}^{T}A_{1}^{T}&=&\frac{1}{2}\begin{pmatrix}
    \cos(2\theta)&\sin(2\theta) \\
    -\sin(2\theta)& \cos(2\theta)
\end{pmatrix},\nonumber\\
A_{1}^{T}A_{2}^{T}&=&\frac{1}{2}\begin{pmatrix}
    -\sin(2\theta)&\cos(2\theta) \\
    -\cos(2\theta)& -\sin(2\theta)
\end{pmatrix}\nonumber\\
A_{1}^{T}A_{3}^{T}&=&\frac{1}{2}\begin{pmatrix}
    \cos(\theta)&-\sin(\theta) \\
    -\sin(\theta)& -\cos(\theta)
\end{pmatrix},\nonumber\\
A_{1}^{T}A_{4}^{T}&=&\frac{1}{2}\begin{pmatrix}
    \sin(\theta)&\cos(\theta) \\
    \cos(\theta)& -\sin(\theta)
\end{pmatrix}\nonumber\\
A_{2}^{T}A_{1}^{T}&=&\frac{1}{2}\begin{pmatrix}
    -\sin(2\theta)&\cos(2\theta) \\
    -\cos(2\theta)& -\sin(2\theta)
\end{pmatrix}\nonumber\\
A_{2}^{T}A_{2}^{T}&=&\frac{1}{2}\begin{pmatrix}
    -\cos(2\theta)&-\sin(2\theta) \\
    \sin(2\theta)& -\cos(2\theta)
\end{pmatrix}\nonumber\\
A_{2}^{T}A_{3}^{T}&=&\frac{1}{2}\begin{pmatrix}
    -\sin(\theta)&-\cos(\theta) \\
    -\cos(\theta)& \sin(\theta)
\end{pmatrix}\nonumber\\
A_{2}^{T}A_{4}^{T}&=&\frac{1}{2}\begin{pmatrix}
    \cos(\theta)&-\sin(\theta) \\
    -\sin(\theta)& -\cos(\theta)
\end{pmatrix}\nonumber\\
M_{3}^{T}M_{1}^{T}&=&\frac{1}{2}\begin{pmatrix}
    \cos(\theta)&\sin(\theta) \\
    \sin(\theta)& -\cos(\theta)
\end{pmatrix}\nonumber\\
A_{3}^{T}A_{2}^{T}&=&\frac{1}{2}\begin{pmatrix}
    -\sin(\theta)&\cos(\theta) \\
    \cos(\theta)& \sin(\theta)
\end{pmatrix}\nonumber\\
A_{3}^{T}A_{3}^{T}&=&\frac{1}{2}\begin{pmatrix}
    1&0 \\
    0& 1
\end{pmatrix}\nonumber\\
A_{3}^{T}A_{4}^{T}&=&\frac{1}{2}\begin{pmatrix}
    0&1\\
    -1& 0
\end{pmatrix}\nonumber\\
A_{4}^{T}A_{1}^{T}&=&\frac{1}{2}\begin{pmatrix}
    -\sin(\theta)&\cos(\theta)\\
    \cos(\theta)& \sin(\theta)
\end{pmatrix}\nonumber\\
A_{4}^{T}A_{2}^{T}&=&\frac{1}{2}\begin{pmatrix}
    -\cos(\theta)&-\sin(\theta)\\
    -\sin(\theta)& \cos(\theta)
\end{pmatrix}\nonumber\\
A_{4}^{T}A_{3}^{T}&=&\frac{1}{2}\begin{pmatrix}
    0&-1\\
   1& 0
\end{pmatrix}\nonumber\\
A_{4}^{T}A_{4}^{T}&=&\frac{1}{2}\begin{pmatrix}
    1&0\\
   0& 1
\end{pmatrix}\nonumber
\end{eqnarray}
Then,  by using these last matrices and  the equations  \eqref{InvRot1}-\eqref{InvRot4} we get the following states 
\begin{eqnarray}
\ket{\phi_{1}}&=&\ket{\psi}\otimes\ket{V_{1}} \nonumber\\
&=&\ket{V_{1}}\frac{1}{2}\left( \left(\cos(2\theta) \alpha_{1}+\sin(2\theta) \alpha_{2}\right)\ket{0}+\left(-\sin(2\theta) \alpha_{1}+\cos(2\theta) \alpha_{2}\right)\ket{1} \right)+\nonumber\\
& &+
\ket{V_{2}}\frac{1}{2}\left( \left(-\sin(2\theta) \alpha_{1}+\cos(2\theta) \alpha_{2}\right)\ket{0}-\left(\cos (2\theta) \alpha_{1}+\sin(2\theta) \alpha_{2}\right)\ket{1} \right)+\nonumber\\
& &+\ket{V_{3}}\frac{1}{2}\left( \left(\cos(\theta) \alpha_{1}-\sin(\theta) \alpha_{2}\right)\ket{0}
-\left(\sin (\theta) \alpha_{1}+\cos(\theta) \alpha_{2}\right)\ket{1} \right)+\nonumber\\
& &+\ket{V_{4}}\frac{1}{2}\left( \left(\sin(\theta) \alpha_{1}+\cos(\theta) \alpha_{2}\right)\ket{0}
+\left(\cos (\theta) \alpha_{1}-\sin(\theta) \alpha_{2}\right)\ket{1} \right),\label{TelRot1} \\
\ket{\phi_{2}}&=&\ket{\psi}\otimes\ket{V_{2}} \nonumber\\
&=&\ket{V_{1}}\frac{1}{2}\left( \left(-\sin(2\theta) \alpha_{1}+\cos(2\theta) \alpha_{2}\right)\ket{0}-\left(\cos(2\theta) \alpha_{1}+\sin(2\theta) \alpha_{2}\right)\ket{1} \right)+\nonumber\\
& &+
\ket{V_{2}}\frac{1}{2}\left( -\left(\cos(2\theta) \alpha_{1}+\sin(2\theta) \alpha_{2}\right)\ket{0}+\left(\sin (2\theta) \alpha_{1}-\cos(2\theta) \alpha_{2}\right)\ket{1} \right)+\nonumber\\
& &+\ket{V_{3}}\frac{1}{2}\left( -\left(\sin(\theta) \alpha_{1}+\cos(\theta) \alpha_{2}\right)\ket{0}+
\left(-\cos (\theta) \alpha_{1}+\sin(\theta) \alpha_{2}\right)\ket{1} \right)+\nonumber\\
& &+\ket{V_{4}}\frac{1}{2}\left( \left(\cos(\theta) \alpha_{1}-\sin(\theta) \alpha_{2}\right)\ket{0}
-\left(\sin (\theta) \alpha_{1}+\cos(\theta) \alpha_{2}\right)\ket{1} \right),\label{TelRot2} \\
\ket{\phi_{3}}&=&\ket{\psi}\otimes\ket{V_{3}} \nonumber\\
&=&\ket{V_{1}}\frac{1}{2}\left( \left(\cos(\theta) \alpha_{1}+\sin(\theta) \alpha_{2}\right)\ket{0}+\left(\sin(\theta) \alpha_{1}-\cos(\theta) \alpha_{2}\right)\ket{1} \right)+\nonumber\\
& &+
\ket{V_{2}}\frac{1}{2}\left( \left(-\sin(\theta) \alpha_{1}+
\cos(\theta) \alpha_{2}\right)\ket{0}+
\left(\cos (\theta) \alpha_{1}+\sin(\theta) \alpha_{2}\right)\ket{1} \right)+\nonumber\\
& &+\ket{V_{3}}\frac{1}{2}\left(\alpha_{1} \ket{0}+\alpha_{2}\ket{1} \right)+\ket{V_{4}}\frac{1}{2}\left( \alpha_{2}\ket{0}-
\alpha_{1}\ket{1} \right),\label{TelRot3} \\
\ket{\phi_{4}}&=&\ket{\psi}\otimes\ket{V_{4}} \nonumber\\
&=&\ket{V_{1}}\frac{1}{2}\left( \left(-\sin(\theta) \alpha_{1}+\cos(\theta) \alpha_{2}\right)\ket{0}+\left(\cos(\theta) \alpha_{1}+\sin(\theta) \alpha_{2}\right)\ket{1} \right)+\nonumber\\
& &+
\ket{V_{2}}\frac{1}{2}\left( -\left(\cos(\theta) \alpha_{1}+
\sin(\theta) \alpha_{2}\right)\ket{0}+
\left(-\sin (\theta) \alpha_{1}+\cos(\theta) \alpha_{2}\right)\ket{1} \right)+\nonumber\\
& &+\ket{V_{3}}\frac{1}{2}\left(-\alpha_{2} \ket{0}+\alpha_{1}\ket{1} \right)+\ket{V_{4}}\frac{1}{2}\left(  \alpha_{1}\ket{0}+
\alpha_{2}\ket{1} \right).\label{TelRot4} 
\end{eqnarray}

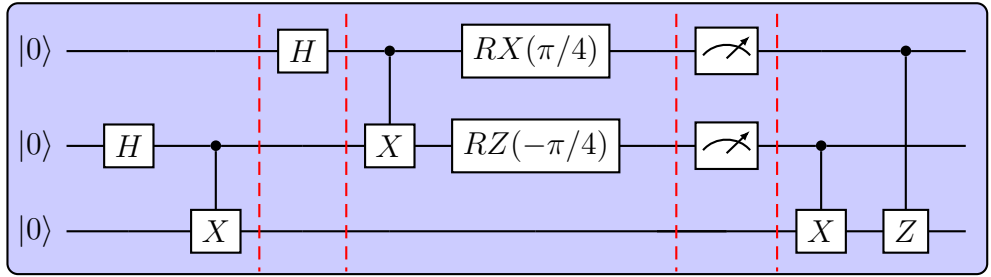
\begin{figure}[hbt!]
\centering
\begin{quantikz}%[color=black,background color=yellow]
\gategroup[wires=3,steps=12,style={rounded corners,fill=blue!20}, background]{}
&\lstick{$|{0}\rangle$} & \qw & \qw
 \slice{} & \gate{H} \slice{} & \ctrl{1} & \gate{RX(\pi / 4)} & \qw \slice{} & \meter{} \slice{} & \qw & \ctrl{2} & \qw
\\
&\lstick{$|{0}\rangle$}  & \gate{H} & \ctrl{1} & \qw & \gate{X} & \gate{RZ(- \pi / 4)} &\qw  & \meter{} & \ctrl{1} & \qw & \qw
\\
&\lstick{$|{0}\rangle$}  & \qw & \gate{X} & \qw & \qw & \qw & \qw & \qw\qw & \gate{X} & \gate{Z} & \qw
\end{quantikz}
\caption{Teleportation with   \eqref{TelRot1}- \eqref{TelRot4}. }
    \label{fig:TelRot}
\end{figure}

In the Figure \ref{fig:TelRot} we present the quantum circuit  for  teleportation with the states  \eqref{TelRot1}- \eqref{TelRot4} and in the  Listing \ref{code:TelRot} we present the asociated code in  Python language using IBM's Qiskit library.

\FloatBarrier

\begin{lstlisting}[language=Python, caption=Code  for  teleportation with the states   \eqref{TelRot1}- \eqref{TelRot4}. , label={code:TelRot}  ]
from qiskit import QuantumRegister, ClassicalRegister, 
from numpy import pi
import math

theta = pi / 4  

qreg_q = QuantumRegister(3, 'q')
creg_c = ClassicalRegister(2, 'c')
qc = QuantumCircuit(qreg_q, creg_c)

qc.x(0) 

qc.h(1) 
qc.cx(1, 2)  

qc.cx(0, 1) 
qc.rx(2 * theta, 0)  
qc.rz(2 * theta, 1)  

qc.measure([0, 1], [0, 1])

qc.cx(qreg_q[1], qreg_q[2])
qc.cz(qreg_q[0], qreg_q[2])
\end{lstlisting}
\FloatBarrier

The algorithm given in Listing \ref{code:TelRot} works as follows:
\begin{enumerate}
\item Entangle the Alice and Bob qubits (is the same entanglement we see in basic Bell states).
\item Prepare the state we want to teleport.
\item In the original algorithm, we apply the Bell Measurement, which means we will collapse the target qubit into one of the Bell states. In the generalized algorithm, we must adapt this circuit to the basis we are working with. 
\item Apply the classic measurement of the target and Alice qubits.
\item Apply the corrections using Controlled NOT and Controlled Z gates. This parts works for every basis.
\end{enumerate}

\section{Example 3}
\label{SecExample3}

For example, using the matrices  
\begin{eqnarray}
A_{1}&=&\frac{1}{\sqrt{2\cosh(2\theta)}}\begin{pmatrix}
\cosh \theta &\sinh \theta \\
\sinh \theta &\cosh\theta  
\end{pmatrix}, \nonumber\\
A_{2}&=&\frac{1}{\sqrt{2\cosh(2\theta)}}\begin{pmatrix}
\sinh\theta  &-\cosh \theta \\
-\cosh\theta &\sinh\theta
\end{pmatrix},\nonumber\\ 
 A_{3}&=&\frac{1}{\sqrt{2}}\begin{pmatrix}
1&0\\
0&-1
\end{pmatrix}, \nonumber\\
A_{4}&=&\frac{1}{\sqrt{2}}\begin{pmatrix}
0&1\\
-1&0
\end{pmatrix}, \nonumber
\end{eqnarray}
the Bell's base can be constructed 
\begin{eqnarray}
\ket{V_{1}}&=&\frac{1}{\sqrt{2\cosh(2\theta)}}\left(\cosh \theta \ket{00}+\sinh \theta \ket{01}+
\sinh \theta \ket{10}+\cosh \theta\ket{11}\right),\label{sbHyper1} \\
\ket{V_{2}}&=&\frac{1}{\sqrt{2\cosh(2\theta)}}\left(\sinh \theta \ket{00}-\cosh \theta \ket{01}-
\cosh \theta\ket{10}+\sinh \theta \ket{11}\right), \label{sbHyper2}  \\
\ket{V_{3}}&=&\frac{1}{\sqrt{2}}\left(\ket{00}-\ket{11}\right),  \label{sbHyper3} \\
\ket{V_{4}}&=&\frac{1}{\sqrt{2}}\left(\ket{01}-\ket{10}\right).  \label{sbHyper4} 
\end{eqnarray}

\begin{figure}[hbt!]
\centering
\begin{quantikz}
\gategroup[wires=2,steps=7,style={rounded corners,fill=blue!20}, background]{} 
&\lstick{$|{0}\rangle$} & & \gate{R_y\left(x'\right)} & \ctrl{1} & \qw & \qw \\
&\lstick{$|{0}\rangle$} & & \qw & \targ{} & \gate{R_y\left(y'\right)} & \qw
\end{quantikz}
\caption{Quantum circuit for the state \eqref{sbHyper1}. }
\label{fig:BellHyper1}
\end{figure}
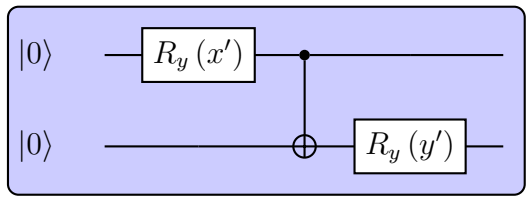

In the figure \ref{fig:BellHyper1} we show  the quantum circuit for the  state \eqref{sbHyper1}  and in the Listing \ref{code:BellHyper1} we present their code in  Python language using IBM's Qiskit library.

\begin{lstlisting}[language=Python, caption=Code for the state \eqref{sbHyper1}. , label={code:BellHyper1}  ]
from qiskit import QuantumCircuit, Aer, execute
from math import pi, cosh, sinh, sqrt, atan2, asin
import numpy as np

def create_circuit(theta):
    qc = QuantumCircuit(2)

    angle_ry1 = 2 * atan2(sinh(theta), cosh(theta))
    angle_ry2 = 2 * asin(sqrt(sinh(theta)**2 / (cosh(theta)**2 + sinh(theta)**2)))

    qc.ry(angle_ry1, 0)

    qc.cx(0, 1)

    qc.ry(angle_ry2, 1)

    return qc

theta = pi / 4
\end{lstlisting}

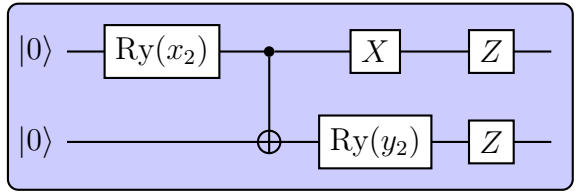
\begin{figure}[hbt!]
\centering
\begin{quantikz}
\gategroup[wires=2,steps=7,style={rounded corners,fill=blue!20}, background]{} 
&\lstick{$|{0}\rangle$} &   \gate{\text{Ry}(x_2)} & \ctrl{1} & \gate{X} & \gate{Z} & \qw \\
&\lstick{$|{0}\rangle$} &  \qw & \targ{} & \gate{\text{Ry}(y_2)} & \gate{Z} & \qw
\end{quantikz}
\caption{Quantum circuit for the state \eqref{sbHyper2}. }
\label{fig:BellHyper2}
\end{figure}

In the figure \ref{fig:BellHyper2} we show  the quantum circuit for the  state \eqref{sbHyper2} and in the Listing \ref{code:BellHyper2} we present their code in  Python language using IBM's Qiskit library.

\begin{lstlisting}[language=Python, caption=Code for the state \eqref{sbHyper2}. , label={code:BellHyper2}  ]
from qiskit import QuantumCircuit, Aer, execute
from math import pi,  cosh, sinh, sqrt
import numpy as np

def create_circuit(theta):
    qc = QuantumCircuit(2)

    ry_angle_1 = 2 * asin(sinh(theta) / sqrt(2 * cosh(2 * theta)))
    ry_angle_2 = 2 * asin(sin(theta) / sqrt(2 * cosh(2 * theta)))

    qc.ry(ry_angle_1, 0)

    qc.cx(0, 1)

    qc.ry(ry_angle_2, 1)

    qc.z(1)
    qc.x(0)
    qc.z(0)

    return qc

theta = pi / 4
\end{lstlisting}

Notice that the states  \eqref{sbHyper3}  and  \eqref{sbHyper4}   are the states   \eqref{sb2} and \eqref{sb4}, the quantum circuits and codes are on the Section \ref{SecBell}.\\

In addition, in this case we have the matrix
\begin{eqnarray}
T=\frac{1}{\sqrt{2\cosh(2\theta)}}\begin{pmatrix}
\cosh \theta &\sinh\theta&\sinh \theta &\cosh \theta\\
\sinh \theta &-\cosh \theta &-\cosh\theta&\sinh \theta\\
\sqrt{\cosh(2\theta)}&0&0&-\sqrt{\cosh(2\theta)}\\
0&\sqrt{\cosh(2\theta)}&-\sqrt{\cosh(2\theta)}&0
\end{pmatrix}
\end{eqnarray}
and 
\begin{eqnarray}
T^{-1}=T^{\dagger}=T^{T}=\frac{1}{\sqrt{2\cosh(2\theta)}}\begin{pmatrix}
\cosh \theta &\sinh \theta & \sqrt{\cosh(2\theta)}& 0\\
\sinh \theta &-\cosh \theta &0&\sqrt{\cosh(2\theta)}\\
\sinh \theta &-\cosh
\theta &0&-\sqrt{\cosh(2\theta)}\\
\cosh \theta &\sinh \theta &-\sqrt{\cosh(2\theta)}&0
\end{pmatrix}
\end{eqnarray}
Then 
\begin{eqnarray}
\ket{00}&=&\frac{1}{\sqrt{2\cosh(2\theta)}}
\left(\cosh \theta \ket{V_{1}}+\sinh \theta \ket{V_{2}}+ \sqrt{\cosh(2\theta)}\ket{V_{3}}\right),\label{InvHyper1} \\
\ket{01}&=&\frac{1}{\sqrt{2\cosh(2\theta)}}\left(\sinh \theta \ket{V_{1}}-\cosh \theta\ket{V_{2}}+ \sqrt{\cosh(2\theta)}\ket{V_{4}}\right),\label{InvHyper2} \\
\ket{10}&=&\frac{1}{\sqrt{2\cosh(2\theta)}}\left(\sinh \theta \ket{V_{1}}-\cosh \theta\ket{V_{2}}-\sqrt{\cosh(2\theta)}\ket{V_{4}}\right),
\label{InvHyper3}\\
\ket{11}&=&\frac{1}{\sqrt{2\cosh(2\theta)}}\left(\cosh \theta \ket{V_{1}}+\sinh \theta \ket{V_{2}}+\sqrt{\cosh(2\theta)}\ket{V_{3}}\right). \label{InvHyper4}
\end{eqnarray}

In addition we have 
\begin{eqnarray}
A_{1}^{T}A_{1}^{T}&=&\frac{1}{2}\begin{pmatrix}
    1&\tanh(2\theta) \\
    \tanh(2\theta)& 1
\end{pmatrix},\nonumber\\
A_{1}^{T}A_{2}^{T}&=&\frac{1}{2\cosh(2\theta)}\begin{pmatrix}
    0&-1 \\
    -1& 0
\end{pmatrix}\nonumber\\
A_{1}^{T}A_{3}^{T}&=&\frac{1}{2\sqrt{\cosh(2\theta)}}\begin{pmatrix}
    \cosh(\theta)&-\sinh(\theta) \\
    \sinh(\theta)& -\cosh(\theta)
\end{pmatrix},\nonumber\\
A_{1}^{T}A_{4}^{T}&=&\frac{1}{2\sqrt{\cosh(2\theta)}}\begin{pmatrix}
    \sinh\theta &-\cosh\theta  \\
    \cosh\theta & -\sinh\theta
\end{pmatrix}\nonumber\\
A_{2}^{T}A_{1}^{T}&=&\frac{1}{2\cosh(2\theta)}\begin{pmatrix}
    0&-1\\
    -1& 0
\end{pmatrix}\nonumber\\
A_{2}^{T}A_{2}^{T}&=&\frac{1}{2}\begin{pmatrix}
    1&-\tanh(2\theta) \\
    -\tanh(2\theta)& 1
\end{pmatrix}\nonumber\\
A_{2}^{T}A_{3}^{T}&=&\frac{1}{2\sqrt{\cosh(2\theta)}}\begin{pmatrix}
    \sinh(\theta)&\cosh(\theta) \\
    -\cosh(\theta)& -\sinh(\theta)
\end{pmatrix}\nonumber\\
A_{2}^{T}A_{4}^{T}&=&\frac{1}{2\sqrt{\cosh(2\theta)}}\begin{pmatrix}
    -\cosh(\theta)&-\sinh(\theta) \\
    \sinh(\theta)& \cosh(\theta)
\end{pmatrix}\nonumber\\
A_{3}^{T}A_{1}^{T}&=&\frac{1}{2\sqrt{\cosh(2\theta)}}\begin{pmatrix}
    \cosh(\theta)&\sinh(\theta) \\
    -\sinh(\theta)& -\cosh(\theta)
\end{pmatrix}\nonumber\\
A_{3}^{T}A_{2}^{T}&=&\frac{1}{2\sqrt{\cosh(2\theta)}}\begin{pmatrix}
    \sinh(\theta)&-\cosh(\theta) \\
    \cosh(\theta)& -\sinh(\theta)
\end{pmatrix}\nonumber\\
A_{3}^{T}A_{3}^{T}&=&\frac{1}{2}\begin{pmatrix}
    1&0 \\
    0& 1
\end{pmatrix}\nonumber\\
A_{3}^{T}A_{4}^{T}&=&\frac{1}{2}\begin{pmatrix}
    0&-1\\
    -1& 0
\end{pmatrix}\nonumber\\
A_{4}^{T}A_{1}^{T}&=&\frac{1}{2\sqrt{\cosh(2\theta)}}\begin{pmatrix}
    -\sinh(\theta)&-\cosh(\theta)\\
    \cosh(\theta)& \sinh(\theta)
\end{pmatrix}\nonumber\\
A_{4}^{T}A_{2}^{T}&=&\frac{1}{2\sqrt{\cosh(2\theta)}}\begin{pmatrix}
    \cosh(\theta)&-\sinh(\theta)\\
    \sinh(\theta)& -\cosh(\theta)
\end{pmatrix}\nonumber\\
A_{4}^{T}A_{3}^{T}&=&\frac{1}{2}\begin{pmatrix}
    0&1\\
   1& 0
\end{pmatrix}\nonumber\\
A_{4}^{T}A_{4}^{T}&=&\frac{1}{2}\begin{pmatrix}
    -1&0\\
   0& -1
\end{pmatrix}\nonumber
\end{eqnarray}

Hence,  by using these last matrices and  the equations  \eqref{InvHyper1}-\eqref{InvHyper4} we obtain the following states 

\begin{eqnarray}
\ket{\phi_{1}}&=&\ket{\psi}\otimes\ket{V_{1}} \nonumber\\
&=&\ket{V_{1}}\frac{1}{2}\left(  \left(\alpha_{1}+\tanh(2\theta)\alpha_{2}\right)\ket{0}+\left(\tanh(2\theta) \alpha_{1}+ \alpha_{2}\right)\ket{1} \right)+\nonumber\\
& &+
\ket{V_{2}}\frac{(-1)}{2\cosh(2\theta)}\left(  \alpha_{2}\ket{0}+\alpha_{1}\ket{1} \right)+\nonumber\\
& &+\ket{V_{3}}\frac{1}{2\sqrt{\cosh(2\theta)}}\left( \left(\cosh(\theta) \alpha_{1}-\sinh(\theta) \alpha_{2}\right)\ket{0}
+\left(\sinh (\theta) \alpha_{1}-\cosh(\theta) \alpha_{2}\right)\ket{1} \right)+\nonumber\\
& &+\ket{V_{4}}\frac{1}{2\sqrt{\cosh(2\theta)}}\bigg( \left(\sinh(\theta) \alpha_{1}
-\cosh(\theta) \alpha_{2}\right)\ket{0}+\nonumber\\
& &+\left(\cosh (\theta) \alpha_{1}-\sinh(\theta) \alpha_{2}\right)\ket{1} \bigg),\label{TelHyper1}\\
\ket{\phi_{2}}&=&\ket{\psi}\otimes\ket{V_{2}} \nonumber\\
&=&\ket{V_{1}}\frac{(-1)}{2\cosh(2\theta)}\left(  \alpha_{2}\ket{0}+ \alpha_{1}\ket{1} \right)+\nonumber\\
& &+
\ket{V_{2}}\frac{1}{2}\left( \left(\alpha_{1}-\tanh(2\theta) \alpha_{2}\right)\ket{0}+\left(-\tanh (2\theta) \alpha_{1}+\alpha_{2}\right)\ket{1} \right)+\nonumber\\
& &+\ket{V_{3}}\frac{1}{2\sqrt{\cosh(2\theta)}}\left( \left(\sinh(\theta) \alpha_{1}+\cosh(\theta) \alpha_{2}\right)\ket{0}-\left(\cosh (\theta) \alpha_{1}+\sinh(\theta) \alpha_{2}\right)\ket{1} \right)+\nonumber\\
& &+\ket{V_{4}}\frac{1}{2\sqrt{\cosh(2\theta)}}\bigg( -\left(\cosh(\theta) \alpha_{1}+\sinh(\theta) \alpha_{2}\right)\ket{0}+\nonumber\\
& &+\left(\sinh (\theta) \alpha_{1}+\cosh(\theta) \alpha_{2}\right)\ket{1} \bigg),\label{TelHyper2}\\
\ket{\phi_{3}}&=&\ket{\psi}\otimes\ket{V_{3}} \nonumber\\
&=&\ket{V_{1}}\frac{1}{2\sqrt{\cosh(2\theta)}}\left( \left(\sinh(\theta) \alpha_{1}-\cosh(\theta) \alpha_{2}\right)\ket{0}-\left(\sinh(\theta) \alpha_{1}+\cosh(\theta) \alpha_{2}\right)\ket{1} \right)+\nonumber\\
& &+
\ket{V_{2}}\frac{1}{2\sqrt{\cosh(2\theta)}}\left( \left(\sinh(\theta) \alpha_{1}-
\cosh(\theta) \alpha_{2}\right)\ket{0}+
\left(\cosh (\theta) \alpha_{1}-\sinh(\theta) \alpha_{2}\right)\ket{1} \right)+\nonumber\\
& &+\ket{V_{3}}\frac{1}{2}\left(\alpha_{1} \ket{0}+\alpha_{2}\ket{1} \right)-\ket{V_{4}}\frac{1}{2}\left( \alpha_{2}\ket{0}+
\alpha_{1}\ket{1} \right), \label{TelHyper3}\\
\ket{\phi_{4}}&=&\ket{\psi}\otimes\ket{V_{4}} \nonumber\\
&=&\ket{V_{1}}\frac{1}{2\sqrt{\cosh(2\theta)}}\left( -\left(\sinh(\theta) \alpha_{1}+\cosh(\theta) \alpha_{2}\right)\ket{0}+\left(\cosh(\theta) \alpha_{1}+\sinh(\theta) \alpha_{2}\right)\ket{1} \right)+\nonumber\\
& &+
\ket{V_{2}}\frac{1}{2\sqrt{\cosh(2\theta)}}\left( \left(\cosh(\theta) \alpha_{1}-
\sinh(\theta) \alpha_{2}\right)\ket{0}+
\left(\sinh (\theta) \alpha_{1}-\cosh(\theta) \alpha_{2}\right)\ket{1} \right)+\nonumber\\
& &+\ket{V_{3}}\frac{1}{2}\left(\alpha_{2} \ket{0}+\alpha_{1}\ket{1} \right)-\ket{V_{4}}\frac{1}{2}\left(  \alpha_{1}\ket{0}+
\alpha_{2}\ket{1} \right).\label{TelHyper4}
\end{eqnarray}

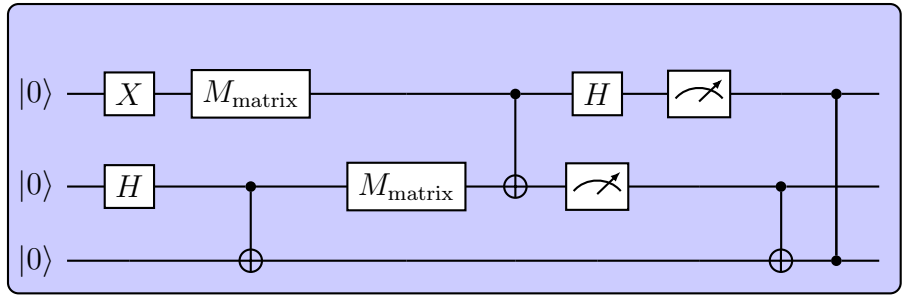
\begin{figure}[hbt!]
\centering
\begin{quantikz}
    \gategroup[wires=4,steps=11,style={rounded corners,fill=blue!20}, background]{}  \\
    &\lstick{$|{0}\rangle$} & \gate{X} & \gate{M_{\text{matrix}}} & \qw &\ctrl{1} & \gate{H} & \meter{} & \qw & \ctrl{2} &\qw \\ 
    &\lstick{$|{0}\rangle$}  & \gate{H} &\ctrl{1} & \gate{M_{\text{matrix}}}  & \targ{} & \meter{} & \qw &\ctrl{1} &\qw &\qw \\
    &\lstick{$|{0}\rangle$} &\qw & \targ{} & \qw & \qw & \qw & \qw & \targ{} & \ctrl{-2} &\qw
\end{quantikz}
\caption{Quantum circuit  for  teleportation with the states   \eqref{TelHyper1}- \eqref{TelHyper4} and   \eqref{TelScale1}- \eqref{TelScale4}}
    \label{fig:TelHyper} 
\end{figure}

In the Figure \ref{fig:TelHyper} we present the quantum circuit  for  teleportation with the states  \eqref{TelHyper1}- \eqref{TelHyper4} and in the  Listing \ref{code:TelHyper} we present the asociated code in  Python language using IBM's Qiskit library.

\FloatBarrier

\begin{lstlisting}[language=Python, caption=Code  for  teleportation with the states   \eqref{TelHyper1}- \eqref{TelHyper4}. , label={code:TelHyper} ]
from qiskit import QuantumRegister, ClassicalRegister, QuantumCircuit,transpile
from qiskit.visualization import plot_histogram
from qiskit.quantum_info import partial_trace, Statevector
from qiskit.quantum_info import DensityMatrix, partial_trace
from qiskit.visualization import plot_state_city
from qiskit_aer import AerSimulator
from numpy import pi
import math
import random


theta = 1.0


M1 = (1/np.sqrt(2 * np.cosh(2 * theta))) * np.array([[np.cosh(theta), np.sinh(theta)], [np.sinh(theta), np.cosh(theta)]])
M2 = (1/np.sqrt(2 * np.cosh(2 * theta))) * np.array([[np.sinh(theta), -np.cosh(theta)], [-np.cosh(theta), np.sinh(theta)]])
M3 = (1/np.sqrt(2)) * np.array([[1, 0], [0, -1]])
M4 = (1/np.sqrt(2)) * np.array([[0, 1], [-1, 0]])


def get_unitary(matrix):
    
    U, S, Vh = np.linalg.svd(matrix)
    return np.dot(U, Vh) 


M1_unitary = get_unitary(M1)
M2_unitary = get_unitary(M2)
M3_unitary = get_unitary(M3)
M4_unitary = get_unitary(M4)


def is_unitary(matrix):
    """Verifica si una matriz es unitaria."""
    identity = np.eye(matrix.shape[0])
    return np.allclose(np.dot(matrix, matrix.conj().T), identity)


for i, M in enumerate([M1_unitary, M2_unitary, M3_unitary, M4_unitary], start=1):
    print(f"M{i} es unitaria:", is_unitary(M))

qreg_q = QuantumRegister(3, 'q')
creg_c = ClassicalRegister(2, 'c')

qc = QuantumCircuit(qreg_q, creg_c)

qc.x(0)


qc.h(1)
qc.cx(1, 2)


matrix_choice = random.choice([M1_unitary, M2_unitary, M3_unitary, M4_unitary])
qc.unitary(matrix_choice, [0, 1], label='M_matrix')  


qc.cx(0, 1)
qc.h(0)


qc.measure([0, 1], [0, 1])

qc.cx(qreg_q[1], qreg_q[2])
qc.cz(qreg_q[0], qreg_q[2])
qc.save_density_matrix(qubits=[2])

print(qc)

simulator = AerSimulator()
circ = transpile(qc, simulator)

state = simulator.run(qc).result().data()['density_matrix']
print(state)

\end{lstlisting}
\FloatBarrier

The algorithm given in Listing \ref{code:TelHyper} works as follows:
\begin{enumerate}
\item Entangle the Alice and Bob qubits (is the same entanglement we see in basic Bell states).
\item Prepare the state we want to teleport.
\item In this example, we seek to collapse the state we want to teleport to  $\ket{V_{1}}, \ket{V_{2}},\ket{V_{3}}, \ket{V_{4}} $  states. The problem is that the matrices in this example are not unitary, so we apply the SVD method in order to get a unitary approximation and after that we can collapse the target qubit to this basis.
\item Apply the classic measurement of the target and Alice qubits.
\item Apply the corrections using Controlled $NOT$ and Controlled $Z$ gates. This parts works for every basis.
\end{enumerate}

In order to obtain a hermitian unitari, in this code we employed the method given in \cite{Xu}.

\section{Example 4}
\label{SecExample4}

For example, using the matrices  
\begin{eqnarray}
A_{1}&=&\frac{1}{\sqrt{1+\lambda^{2}}}\begin{pmatrix}
\lambda &0 \\
0&1
\end{pmatrix}, \nonumber\\
A_{2}&=&\frac{1}{\sqrt{1+\lambda^{2}}}\begin{pmatrix}
1 &0\\
0 &-\lambda
\end{pmatrix},\nonumber\\ 
 A_{3}&=&\frac{1}{\sqrt{2}}\begin{pmatrix}
0&1\\
-1&0
\end{pmatrix}, \nonumber\\
A_{4}&=&\frac{1}{\sqrt{2}}\begin{pmatrix}
0&1\\
1&0
\end{pmatrix}, \nonumber
\end{eqnarray}
the Bell's base can be   constructed 
\begin{eqnarray}
\ket{V_{1}}&=&\frac{1}{\sqrt{1+\lambda^{2}}}\left(\lambda \ket{00}+ \ket{11}\right),\label{sbEscale1} \\
\ket{V_{2}}&=&\frac{1}{\sqrt{1+\lambda^{2}}}\left( \ket{00}-\lambda  \ket{11}\right),\label{sbEscale2} \\
\ket{V_{3}}&=&\frac{1}{\sqrt{2}}\left(\ket{01}-\ket{10}\right),\label{sbEscale3}\\
\ket{V_{4}}&=&\frac{1}{\sqrt{2}}\left(\ket{01}+\ket{10}\right). \label{sbEscale4}
\end{eqnarray}

\begin{figure}[hbt!]
\centering
\begin{quantikz}%[color=black,background color=yellow]
\gategroup[wires=2,steps=8,style={rounded corners,fill=blue!20}, background]{}
&\lstick{$|{0}\rangle$} & \qw& \gate{H} & 
 \gate{RY(2(arctan(1/\alpha))} & \ctrl{1} \slice{} & \meter{} & \qw
\\
&\lstick{$|{0}\rangle$}  & \qw & \qw & \qw & \gate{X} & \meter{} & \qw
\end{quantikz}
\caption{Quantum circuit for the state \eqref{sbEscale1}.  }
    \label{fig:BellEscale1} 
\end{figure}
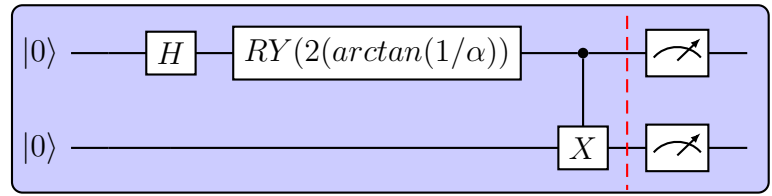

In the figure \ref{fig:BellEscale1} we show  the quantum circuit for the  states \eqref{sbEscale1} and in the Listing \ref{code:BellEscale1} we present their code in  Python language using IBM's Qiskit library.

\FloatBarrier 

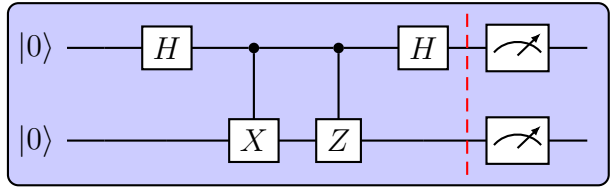
\begin{figure}[hbt!]
\centering
\begin{quantikz}%[color=black,background color=yellow]
\gategroup[wires=2,steps=9,style={rounded corners,fill=blue!20}, background]{}
&\lstick{$|{0}\rangle$} & \qw& \gate{H} & 
 \ctrl{1} & \ctrl{1} & \gate{H} \slice{} & \meter{} & \qw
\\
&\lstick{$|{0}\rangle$}  & \qw & \qw & \gate{X} & \gate{Z} & \qw & \meter{} & \qw
\end{quantikz}
\caption{Quantum circuit for the state \eqref{sbEscale2}. }
    \label{fig:BellEscale2}
\end{figure}

\begin{lstlisting}[language=Python, caption=Code for the state \eqref{sbEscale1}. , label={code:BellEscale1} ]
import numpy as np
import matplotlib.pyplot as plt
from qiskit import QuantumCircuit, QuantumRegister
import qiskit.quantum_info as qi
from math import atan, sqrt, pi

alpha = 2.0
theta = 2 * atan(1 / alpha)

qreg_q = QuantumRegister(2, 'q')
creg_c = ClassicalRegister(4, 'c')
circuit = QuantumCircuit(qreg_q, creg_c)

circuit.h(qreg_q[0])
circuit.barrier(qreg_q[0], qreg_q[1])
circuit.ry(theta, qreg_q[0])
circuit.cx(qreg_q[0], qreg_q[1])
circuit.barrier(qreg_q[0], qreg_q[1])
circuit.swap(qreg_q[0], qreg_q[1])
\end{lstlisting}
In the figure \ref{fig:BellEscale2} we show  the quantum circuit for the  state \eqref{sbEscale2} and in the Listing \ref{code:BellEscale2} we present their code in  Python language using IBM's Qiskit library.

\FloatBarrier 

\begin{lstlisting}[language=Python, caption=Code for the state \eqref{sbEscale2}. , label={code:BellEscale2}  ]
import numpy as np
import matplotlib.pyplot as plt
from qiskit import QuantumCircuit, QuantumRegister, assemble
import qiskit.quantum_info as qi
from math import atan, sqrt, pi
from qiskit_aer import AerSimulator, UnitarySimulator, StatevectorSimulator

alpha = 2.0
maxFact = 1 / sqrt(1 + alpha**2)

qreg_q = QuantumRegister(2, 'q')
creg_c = ClassicalRegister(4, 'c')
circuit = QuantumCircuit(qreg_q, creg_c)

circuit.h(qreg_q[0])
circuit.barrier(qreg_q[0], qreg_q[1])
circuit.cx(qreg_q[0], qreg_q[1])
circuit.cz(qreg_q[1], qreg_q[0])
circuit.h(qreg_q[0])
circuit.barrier(qreg_q[0], qreg_q[1])
circuit.swap(qreg_q[0], qreg_q[1])
\end{lstlisting}

Notice that the states  \eqref{sbEscale3}  and  \eqref{sbEscale4} are the states   \eqref{sb3} and \eqref{sb4}, the quantum circuits and codes are on the Section \ref{SecBell}.\\

In addition, in this case we have the matrix

\begin{eqnarray}
T=\frac{1}{\sqrt{2\left(1+\lambda^{2}\right)}}\begin{pmatrix}
\sqrt{2}\lambda &0&0 &\sqrt{2}\\
\sqrt{2} &0&0&-\sqrt{2}\lambda\\
0&\sqrt{1+\lambda^{2}}&-\sqrt{1+\lambda^{2}}&0\\
0&\sqrt{1+\lambda^{2}}&\sqrt{1+\lambda^{2}}&0
\end{pmatrix}
\end{eqnarray}
and 
\begin{eqnarray}
T^{-1}=T^{\dagger}=T^{T}=\frac{1}{\sqrt{2\left(1+\lambda^{2}\right)}}
\begin{pmatrix}
\sqrt{2}\lambda &\sqrt{2}&0 &0\\
0 &0&\sqrt{1+\lambda^{2}}&\sqrt{1+\lambda^{2}}\\
0&0&-\sqrt{1+\lambda^{2}}&\sqrt{1+\lambda^{2}}\\
\sqrt{2}&-\sqrt{2}\lambda&0&0
\end{pmatrix}
\end{eqnarray}
Then 
\begin{eqnarray}
\ket{00}&=&\frac{1}{\sqrt{1+\lambda^{2}}}
\left(\lambda \ket{V_{1}}+ \ket{V_{2}}\right),\label{InvScale1}   \\
\ket{01}&=&\frac{1}{\sqrt{2}}\left(
\ket{V_{3}}+\ket{V_{4}}\right), \label{InvScale2} \\
\ket{10}&=&\frac{1}{\sqrt{2}}\left(-\ket{V_{3}}+\ket{V_{4}}\right),\label{InvScale3}\\
\ket{11}&=&\frac{1}{\sqrt{1+\lambda^{2}}}\left( \ket{V_{1}}
-\lambda \ket{V_{2}}\right). \label{InvScale4}
\end{eqnarray}

In addition we have
\begin{eqnarray}
A_{1}^{T}A_{1}^{*T}&=&\frac{1}{1+\lambda^{2}}\begin{pmatrix}
    \lambda^{2}&0\\
    0& 1
\end{pmatrix},\nonumber\\
A_{1}^{T}A_{2}^{*T}&=&\frac{1}{1+\lambda^{2}}\begin{pmatrix}
    \lambda&0 \\
    0& - \lambda
\end{pmatrix}\nonumber\\
A_{1}^{T}A_{3}^{*T}&=&\frac{1}{\sqrt{2(1+\lambda^{2})}}\begin{pmatrix}
    0&-\lambda \\
    1& 0
\end{pmatrix},\nonumber\\
A_{1}^{T}A_{4}^{*T}&=&\frac{1}{\sqrt{2(1+\lambda^{2})}}\begin{pmatrix}
    0&\lambda \\
    1& 0
\end{pmatrix}\nonumber\\
A_{2}^{T}A_{1}^{*T}&=&\frac{1}{1+\lambda^{2}}\begin{pmatrix}
    \lambda&0\\
    0&  -\lambda
\end{pmatrix}\nonumber\\
A_{2}^{T}A_{2}^{*T}&=&\frac{1}{1+\lambda^{2}}\begin{pmatrix}
    1&0 \\
    0& \lambda^{2}
\end{pmatrix}\nonumber\\
A_{2}^{T}A_{3}^{*T}&=&\frac{1}{\sqrt{2(1+\lambda^{2})}}\begin{pmatrix}
    0&-1 \\
    -\lambda& 0
\end{pmatrix}\nonumber\\
A_{2}^{T}A_{4}^{*T}&=&\frac{1}{\sqrt{2(1+\lambda^{2})}}\begin{pmatrix}
    0&1 \\
    -\lambda& 0
\end{pmatrix}\nonumber\\
A_{3}^{T}A_{1}^{*T}&=&\frac{1}{\sqrt{2(1+\lambda^{2})}}\begin{pmatrix}
    0&-1 \\
    \lambda& 0
\end{pmatrix}\nonumber\\
A_{3}^{T}A_{2}^{*T}&=&\frac{1}{\sqrt{2(1+\lambda^{2})}}\begin{pmatrix}
    0&\lambda \\
    1& 0
\end{pmatrix}\nonumber\\
A_{3}^{T}A_{3}^{*T}&=&\frac{1}{2}\begin{pmatrix}
    -1&0 \\
    0& -1
\end{pmatrix}\nonumber\\
A_{3}^{T}A_{4}^{*T}&=&\frac{1}{2}\begin{pmatrix}
    -1&0\\
    0& 1
\end{pmatrix}\nonumber\\
A_{4}^{T}A_{1}^{*T}&=&\frac{1}{\sqrt{2(1+\lambda^{2})}}\begin{pmatrix}
    0&1\\
    \lambda& 0
\end{pmatrix}\nonumber\\
M_{4}^{T}M_{2}^{*T}&=&\frac{1}{\sqrt{2(1+\lambda^{2})}}\begin{pmatrix}
    0&-\lambda\\
    1& 0
\end{pmatrix}\nonumber\\
A_{4}^{T}A_{3}^{*T}&=&\frac{1}{2}\begin{pmatrix}
    1&0\\
    0& -1
\end{pmatrix}\nonumber\\
A_{4}^{T}A_{4}^{*T}&=&\frac{1}{2}\begin{pmatrix}
    1&0\\
    0& 1
\end{pmatrix}.   \nonumber
\end{eqnarray}

Therefore, by using these last matrices and  the equations  \eqref{InvScale1}-\eqref{InvScale4} we obtain the following states 

\begin{eqnarray}
\ket{\phi_{1}}&=&\ket{\psi}\otimes\ket{V_{1}} \nonumber\\
&=&\ket{V_{1}}\frac{1}{1+\lambda^{2}}\left( \lambda^{2} \alpha_{1}\ket{0}+ \alpha_{2}\ket{1} \right)+\nonumber\\
& &+
\ket{V_{2}}\frac{1}{1+\lambda^{2}}\left(  \lambda\alpha_{1}\ket{0}- \lambda\alpha_{2}\ket{1} \right)+\nonumber\\
& &+\ket{V_{3}}\frac{1}{\sqrt{2(1+\lambda^{2})}}\left(- \lambda\alpha_{2}\ket{0}+\alpha_{1}\ket{1} \right)+\nonumber\\
& &+\ket{V_{4}}\frac{1}{\sqrt{2(1+\lambda^{2})}}\left( \lambda\alpha_{2}\ket{0}+ \alpha_{1}\ket{1} \right)\label{TelScale1}\\
\ket{\phi_{2}}&=&\ket{\psi}\otimes\ket{V_{2}} \nonumber\\
&=&\ket{V_{1}}\frac{1}{1+\lambda^{2}}\left(  \lambda\alpha_{1}\ket{0}- \lambda\alpha_{2}\ket{1} \right)+\nonumber\\\
& &+
\ket{V_{2}}\frac{1}{1+\lambda^{2}}\left(  \alpha_{1}\ket{0}+\lambda^{2}\alpha_{2}\ket{1} \right)+\nonumber\\
& &+\ket{V_{3}}\frac{(-1)}{\sqrt{2(1+\lambda^{2})}}\left(  \alpha_{2}\ket{0}+\lambda\alpha_{1}\ket{1} \right)+\nonumber\\
& &+\ket{V_{4}}\frac{1}{\sqrt{2(1+\lambda^{2})}}\left(  \alpha_{2}\ket{0}- \lambda\alpha_{1}\ket{1} \right)\label{TelScale2}\\
\ket{\phi_{3}}&=&\ket{\psi}\otimes\ket{V_{3}} \nonumber\\
&=&\ket{V_{1}}\frac{1}{\sqrt{2(1+\lambda^{2})}}\left(  -\alpha_{2}\ket{0}+ \lambda\alpha_{1}\ket{1} \right)+\nonumber\\
& &+
\ket{V_{2}}\frac{1}{\sqrt{2(1+\lambda^{2})}}\left(  \lambda\alpha_{2}\ket{0}+\alpha_{1}\ket{1} \right)+\nonumber\\
& &+\ket{V_{3}}\frac{(-1)}{2}\left(  \alpha_{1}\ket{0}+\alpha_{2}\ket{1} \right)+\nonumber\\
& &+\ket{V_{4}}\frac{1}{2}\left( - \alpha_{1}\ket{0} +\alpha_{2}\ket{1} \right)\nonumber\\
\ket{\phi_{4}}&=&\ket{\psi}\otimes\ket{V_{4}} \label{TelScale3} \\
&=&\ket{V_{1}}\frac{1}{\sqrt{2(1+\lambda^{2})}}\left(  \alpha_{2}\ket{0}+ \lambda\alpha_{1}\ket{1} \right)+\nonumber\\
& &+
\ket{V_{2}} \frac{1}{\sqrt{2(1+\lambda^{2})}} \left(  -\alpha_{2}\lambda\ket{0}+ \alpha_{1}\ket{1} \right)+\nonumber\\
& &+\ket{V_{3}}\frac{1}{2}\left(  \alpha_{1}\ket{0}
-\alpha_{2}\ket{1} \right)+\ket{V_{4}}\frac{1}{2}\left(  \alpha_{1}\ket{0}+
\alpha_{2}\ket{1} \right).\label{TelScale4}
\end{eqnarray}

In the Figure  \ref{fig:TelHyper} we present the quantum circuit  for  teleportation with the states  \eqref{TelScale1}- \eqref{TelScale4} and in the  Listing \ref{code:TelScale} we present the asociated code in  Python language using IBM's Qiskit library.

\begin{lstlisting}[language=Python, caption=Code  for  teleportation with the states   \eqref{TelScale1}- \eqref{TelScale4}.  , label={code:TelScale}  ]
import numpy as np
from qiskit import QuantumCircuit
from qiskit.quantum_info import Statevector, Operator
import random

lambda_val = 1


M1 = (1/np.sqrt(1+lambda_val**2)) * np.array([[lambda_val, 0], [0, 1]])
M2 = (1/np.sqrt(1+lambda_val**2)) * np.array([[1, 0], [0, -lambda_val]])
M3 = (1/np.sqrt(2)) * np.array([[0, 1], [-1, 0]])
M4 = (1/np.sqrt(2)) * np.array([[0, 1], [1, 0]])


def get_unitary(matrix):

    U, S, Vh = np.linalg.svd(matrix)
    return np.dot(U, Vh) 



M1_unitary = get_unitary(M1)
M2_unitary = get_unitary(M2)
M3_unitary = get_unitary(M3)
M4_unitary = get_unitary(M4)

def is_unitary(matrix):

    identity = np.eye(matrix.shape[0])
    return np.allclose(np.dot(matrix, matrix.conj().T), identity)


qreg_q = QuantumRegister(3, 'q')
creg_c = ClassicalRegister(2, 'c')

qc = QuantumCircuit(qreg_q, creg_c)


qc.x(0)


qc.h(1)
qc.cx(1, 2)


matrix_choice = random.choice([M1_unitary, M2_unitary, M3_unitary, M4_unitary])
qc.unitary(matrix_choice, [0, 1], label='M_matrix')  


qc.cx(0, 1)
qc.h(0)


qc.measure([0, 1], [0, 1])

qc.cx(qreg_q[1], qreg_q[2])
qc.cz(qreg_q[0], qreg_q[2])
qc.save_density_matrix(qubits=[2])

print(qc)


simulator = AerSimulator()
circ = transpile(qc, simulator)


state = simulator.run(qc).result().data()['density_matrix']
print(state)

\end{lstlisting}
\FloatBarrier

The algorithm given in Listing \ref{code:TelScale} works as follows:
\begin{enumerate}
\item Entangle the Alice and Bob qubits (is the same entanglement we see in basic Bell states).
\item Prepare the state we want to teleport.
\item In this example, we seek to collapse the state we want to teleport to $\ket{V_{1}}, \ket{V_{2}},\ket{V_{3}}, \ket{V_{4}} $ states. The problem is that the matrices in this example are not Unitary, so we apply the SVD method in order to get a unitary approximation and after that we can collapse the target qubit to this basis.
\item Apply the classic measurement of the target and Alice qubits.
\item Apply the corrections using Controlled $NOT$ and Controlled $Z$ gates. This parts works for every basis.
\end{enumerate}

In order to obtain a  unitary marix, in this code we employed the  SVD method given in \cite{Xu}.

\section{Conclusions}
\label{Con}

We proposed  alternative bases of entangled states and studied quantum teleportation with them. Some of these bases depend on a continuous parameter. In addition,  we presented the quantum circuit and code asociated to bases and teportation. \\

We think that these alternative basis of entangled states can be employed in different applications, such as quantun cryptography.

\section*{Acknowledgement}

\end{document}